\documentclass[aps,pra,reprint,superscriptaddress,amsmath,amssymb]{revtex4-2}

\usepackage{graphicx}
\usepackage[driverfallback=dvipdfm]{hyperref}
\usepackage{tabularx}
\usepackage{mathtools}
\usepackage[mediumspace,mediumqspace,squaren]{SIunits}

\usepackage{xcolor}

\newcommand{\ee} {{\textrm e}}
\newcommand{\dd} {\hbox{\textrm d}}
\newcommand{\erf} {\hbox{\textrm erf}}
\newcommand{\erfc} {\hbox{\textrm erfc}}
\newcommand{\Real}{\text{Re}}
\newcommand{\Imag}{\text{Im}}
\newcommand{\Ei}{\text{Ei}}
\newcommand{\sign}{\text{sign}}
\newcommand{\Tr}{\text{Tr}}
\newcommand{\rank}{\text{rank}}
\newcommand{\bra}[1]{\ensuremath{\left\langle #1 \right|}}
\newcommand{\ket}[1]{\ensuremath{\left| #1 \right\rangle}}

\newcommand{\ketbra}[2]{\left| #1 \middle\rangle\middle\langle #2 \right|}
\newcommand{\braMket}[3]{\ensuremath{\left\langle #1 \middle| #2 \middle| #3 \right\rangle}}
\newcommand{\oa}{{\hat{a}}}
\newcommand{\oad}{{\hat{a}^\dag}}
\newcommand{\oA}{{\hat{A}}}
\newcommand{\ob}{{\hat{b}}}
\newcommand{\oB}{{\hat{B}}}
\newcommand{\obd}{{\hat{b}^\dag}}
\renewcommand{\oc}{{\hat{c}}}
\newcommand{\ocd} {{\hat{c}^\dag}}
\newcommand{\od} {{\hat{d}}}
\newcommand{\odd} {{\hat{d}^\dag}}

\newcommand{\os} {{\hat{\sigma}}}

\newcommand{\oG} {{\hat{G}}}
\newcommand{\oH} {{\hat{H}}}
\newcommand{\oK} {{\hat{K}}}
\newcommand{\oKd} {{\hat{K}^\dag}}
\newcommand{\oL} {{\hat{L}}}
\newcommand{\oLd} {{\hat{L}^\dag}}
\newcommand{\on} {{\hat{n}}}
\newcommand{\oP} {{\hat{P}}}
\newcommand{\oQ} {{\hat{Q}}}
\newcommand{\oQd} {{\hat{Q}^\dag}}
\newcommand{\oS} {{\hat{S}}}
\newcommand{\oSd} {{\hat{S}^\dag}}
\newcommand{\oT} {{\hat{T}}}
\newcommand{\oTd} {{\hat{T}^\dag}}
\newcommand{\oU} {{\hat{U}}}
\newcommand{\oV} {{\hat{V}}}
\newcommand{\oD} {{\hat{D}}}
\newcommand{\ox} {{\hat{x}}}
\newcommand{\op} {{\hat{p}}}
\newcommand{\orho} {{\hat{\rho}}}

\begin{document}
\title{Imperfect blockade in Rydberg superatoms}

\author{Valentin Magro} \affiliation{JEIP,  UAR 3573 CNRS, Coll{\`e}ge de France, PSL University, 11, place Marcelin Berthelot, 75231 Paris Cedex 05, France}

\author{S{\'e}bastien Garcia} \affiliation{JEIP,  UAR 3573 CNRS, Coll{\`e}ge de France, PSL University, 11, place Marcelin Berthelot, 75231 Paris Cedex 05, France}

\author{Alexei Ourjoumtsev} \email{Corresponding author: alexei.ourjoumtsev@college-de-france.fr} \affiliation{JEIP,  UAR 3573 CNRS, Coll{\`e}ge de France, PSL University, 11, place Marcelin Berthelot, 75231 Paris Cedex 05, France}
\begin{abstract}
Ensembles of atoms interacting via their Rydberg levels, known as ``superatoms'' for their ability to encode qubits and to emit single photons, attract increasing attention as building blocks for quantum network nodes. Assessing their performance requires an accurate, physically informative and numerically scalable description of interactions in a large and disordered ensemble. We derive such a description from first principles and successfully test it against brute-force numerics and experimental data. This model proves essential to make quantitative predictions about gate fidelities or photon emission efficiencies, and to guide experiments towards large-scale superatom-based systems. 
\end{abstract}

\maketitle

Photon exchange between quantum emitters is a fundamental physical process underlying many applications in quantum technologies. Its efficiency improves with the ratio between the scattering cross-section of the emitter and the characteristic cross-section of the photon. While confining the photon in a waveguide or in a small-volume cavity is the most explored route towards this goal \cite{Haroche2006,Chang2014}, increasing the effective size of the atom emerged as an interesting alternative \cite{Lukin2001,Dudin2012,Firstenberg2016,Kumlin2023,Shao2024}.  It consists in using a ``superatom'' formed by a small atomic ensemble, where the anharmonicity required to isolate a single qubit is provided by strong interactions between Rydberg excitations. 

Building superatom-based quantum nodes requires accurately predicting their imperfections, in particular those related to Rydberg blockade. Existing models often focus on correlations between the emitted photons, insensitive to linear losses, and consider either continuous driving or instantaneous excitation \cite{Gorshkov2011,Bariani2012,Stanojevic2012,Gorshkov2013,Bienias2014,Grankin2015,Grankin2016, Georgakopoulos2018}. Semi-phenomenological models, accounting for dynamical control and non-Markovian interaction-induced qubit decay, successfully fit existing data \cite{Stiesdal2020} but lack universality and thus predictive ability. This makes it difficult to describe the non-Gaussian photonic states produced with these superatoms \cite{Magro2023}, to predict the success rates of superatom-based multi-photon gates \cite{Stolz2022} and, more generally, to draw conclusions about the scalability of this approach.

Here we show that a dynamically driven and imperfectly blockaded superatom allows an accurate, low-dimensional, bottom-up description, scalable towards multi-node quantum networks. Derived from first principles and successfully tested against brute-force simulations and experimental data, it precisely describes the system's dynamics, thus accelerating experimental optimizations \cite{Magro2023} and enabling novel control techniques \cite{Aggarwal2026}.

\begin{figure}
\centering
	\includegraphics[width=85mm]{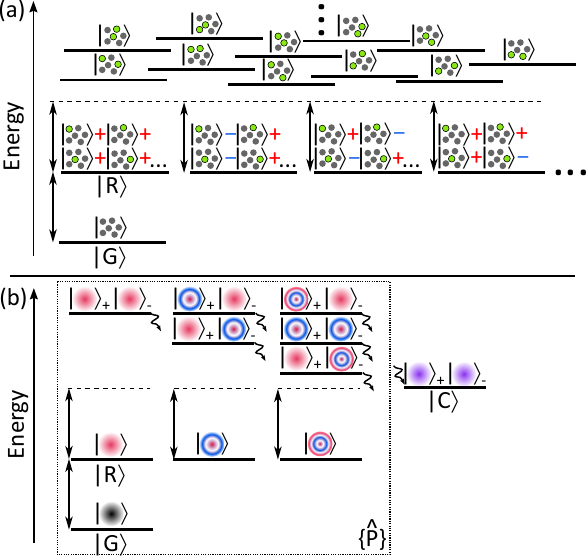}
	\caption{Effective blockade model. (a) In the ``microscopic'' model of Eq.\ref{eq:HDriveDiscrete}, the collective state \ket{G} is coupled to the exchange-symmetric singly-excited state \ket{R}, itself off-resonantly coupled to asymmetric states via the doubly-excited space, where the eigenstates correspond to atomic pairs at well-defined distances. (b) The effective model treats the cloud as a continuous medium and uses the eigenstates of a 3D harmonic oscillator as a basis, including a few of them in the Hamiltonian, and treating the others as a memory-less continuum \ket{C}. The laser coupling between singly- and doubly-excited states preserves the total number of spatial excitations.}
	\label{fig:LevelScheme}
\end{figure}

We consider a disordered ensemble of $N \gg 1$ atoms, resonantly driven between the ground state \ket{g} and a highly-excited Rydberg state \ket{r} where the atoms present strong van-der-Waals interactions (an extension to dipole-dipole interactions is presented in Supplemental Material C.4). The actual interaction potential is more complex at short distance \cite{Weber2017,Sibalic2017,Beguin2013}, but this long-range approximation suffices to describe the dominant effects. Setting $\hbar=1$, the ``microscopic'' Hamiltonian of the system is 
\begin{equation}
\label{eq:HDriveDiscrete}
\oH =  
\frac{\Omega}{2}\sum_{n=1}^N\frac{\os_{gr}^{(n)}+\os_{rg}^{(n)}}{\sqrt{N}} + \frac{1}{2}\sum_{i \neq j}^N \frac{C_6}{|{\bf R}_i-{\bf R}_j|^6}\os_{rr}^{(i)}\os_{rr}^{(j)},
\end{equation}
where $\os_{ij}^{(n)}=|i^{(n)}\rangle\langle j^{(n)}|$ for the $n$-th atom, ${\bf R}_n$ is its position, the coefficient $C_6$ characterizes the van-der-Waals interactions, and the collectively-enhanced Rabi frequency $\Omega$, taken real for simplicity, may depend on time. The Hilbert space of a perfectly Rydberg-blockaded superatom is limited to the ground state $\ket{G}=|g^{(1)},...,g^{(n)}\rangle$ and the exchange-symmetric singly-excited state $\ket{R}=\oSd\ket{G}$ where $\oSd=\sum_{n=1}^N \os_{rg}^{(n)}/\sqrt{N}$. Imperfect blockade, considered here to the lowest order, expands it to $\sim N^2/2$ states containing up to two excitations, see Fig.~\ref{fig:LevelScheme}(a). For typical values $N\sim 10^{3}$, simulations recursively optimizing experimental parameters or describing multiple driven-dissipative and photon-coupled superatoms become impractically slow or resource-intensive. Our goal is to find the smallest possible subspace to accurately describe the system's dynamics, treating the rest of this Hilbert space as a quasi-continuum, cf Fig.~\ref{fig:LevelScheme}(b).

In most experiments, the atomic ensemble is prepared using a dipole trap formed by a single or two crossed Gaussian beams. As it is composed of many disordered atoms, we approximate it with a continuous Gaussian atomic density distribution, with a radius $\sigma$ along $x$ and $y$, and $\sigma_z$ along $z$. For the dimensionless position ${\bf r}=(X/\sigma,Y/\sigma,Z/\sigma_z)$, the normalized distribution is $\mu({\bf r})=\ee^{- r^2/2}/(2\pi)^{3/2}$. We replace discrete atomic operators with continuous ones, satisfying $[\os_{ij}({\bf r}),\os_{ji}({\bf r'})]=\delta^{(3)}({\bf r}-{\bf r'})(\os_{ii}({\bf r})-\os_{jj}({\bf r}))$, and we transfer the Gaussian shape of the atomic density to the Hamiltonian terms, by redefining the collective excitation operator as $\oSd=\int\dd^3{\bf r}\sqrt{\mu({\bf r})}\os_{rg}({\bf r})$. This mapping preserves the global experimentally-accessible properties of the superatom, and the local populations required to describe the blockade. Considering that at most two out of many atoms are excited, we make a lowest-order Holstein-Primakoff approximation $[\oS,\oSd] \approx 1$.

The symmetric singly-excited state $|R\rangle=\oSd|G\rangle=|\psi_{0,0,0}\rangle$ has a Gaussian wavefunction $\psi_{0,0,0}({\bf r})=\sqrt{\mu({\bf r})}$, like the ground state of an isotropic 3D harmonic oscillator. We thus choose the eigenstates $\ket{\psi_{n,l,m}}$ of the latter to form the singly-excited basis. They contain $2n+l$ spatial excitations and co-diagonalize the total angular momentum $\oL^2$ and its projection $\oL_z$ for respective eigenvalues $l(l+1)$ and $m$, allowing us to separate radial and angular degrees of freedom. In Supplemental Material~A, we construct them algebraically to simplify calculations.

The cloud's invariance by rotation around $z$ and by reflection around $xOy$ reduces the populated singly-excited states to $|\psi_{n,l}\rangle=|\psi_{n,2l,0}\rangle$, and to $|\psi_{n}\rangle=|\psi_{n,0,0}\rangle$ for a spherical cloud. They are laser-coupled to doubly-excited states $\ket{\Psi_{n_c,l_c,n_d,l_d,m}}$ where we separate the relative coordinate ${\bf r}_d=({\bf r}_a - {\bf r}_b)/\sqrt{2}$ between the two excitations $a$ and $b$ from their ``center of mass'' ${\bf r}_c=({\bf r}_a + {\bf r}_b)/\sqrt{2}$, unaffected by the interactions. Reducing the basis via symmetries, we show in Supp. Mat.~B that
\begin{equation}
\oSd
= \ketbra{R}{G}+\sum S_{n,l;l_c,n_d,l_d,m} \ketbra{\Psi_{n_c,l_c,n_d,l_d,m}}{\psi_{n,l}}
\end{equation}
where the sum runs over $n_c+n_d+l_c+l_d=n+l$, $|l_c-l_d|\leq l \leq l_c+l_d$, and $0\leq m\leq \min(2l_c,2l_d)$,
\begin{multline}
\label{eq:Psi2exc}
\ket{\Psi_{n_c,l_c,n_d,l_d,m}}
= 
\int\dd{\bf r}_a^3\dd{\bf r}_b^3 \frac{\os_{rg}({\bf r}_a)\os_{rg}({\bf r}_b)}{2\sqrt{1+\delta_{m}}} \\
 \left[ \psi_{n_c,2l_c,-m}({\bf r}_c)\psi_{n_d,2l_d,m}({\bf r}_d)\right.\\
 \left. +\psi_{n_c,2l_c,m}({\bf r}_c)\psi_{n_d,2l_d,-m}({\bf r}_d)\right] \ket{G},
\end{multline}
where $\delta_m$ is the Kronecker symbol, and $S$ is defined via Wigner 3j symbols:
\begin{align}
\label{eq:SLaser}
\nonumber
S_{n,l;l_c,n_d,l_d,m} 
=& \frac{(-1)^{l_c+l_d-l}}{\sqrt{1+\delta_{m}}}\frac{4l+1}{2^{n+l-1 }}\sqrt{\frac{\mathcal{D}_{n_c,l_c}\mathcal{D}_{n_d,l_d}}{\mathcal{D}_{n,l}}} \\
\nonumber
& \left(\begin{array}{ccc}
2l_c & 2l_d & 2l \\ 0 & 0 & 0
\end{array}\right)
\left(\begin{array}{ccc}
2l_c & 2l_d & 2l \\ -m & m & 0
\end{array}\right),\\
\mathcal{D}_{n,l} =& \frac{4l+1}{2^n n!(2n+4l+1)!!}.
\end{align}
For a spherical cloud, this further reduces to
\begin{align}
\label{eq:Psi2excSph}
\nonumber
\oSd
=& \ketbra{R}{G}+\sum S_{n;n_d,l}\ketbra{\Psi_{n_c,n_d,l}}{\psi_{n}},\\
\nonumber
\ket{\Psi_{n_c,n_d,l}}
=& \sum_{m=0}^{2l} \sqrt{\frac{2}{1+\delta_m}} \frac{(-1)^m}{\sqrt{4l+1}}\ket{\Psi_{n_c,l,n_d,l,m}},\\
S_{n;n_d,l} =& \frac{\sqrt{2}}{2^{n}}\sqrt{\frac{\mathcal{D}_{n_c,l}\mathcal{D}_{n_d,l}}{(4l+1)\mathcal{D}_{n,0}}}.
\end{align}
with $n_d+n_c+2l=n$.

To reduce the Hilbert space, we consider the subspace $\{\oP\}$ defined by a projector $\oP$, formed by $\ket{G}$ and a few low-lying singly- and doubly-excited states obeying the selection rules above, with $n+l\leq n_{max}$ chosen as a compromise between the precision and the size of the simulation. In the two-excitation limit, $\{\oP\}$ is stable under the laser-coupling term $(\Omega/2)(\oS+\oSd)$ of $\oH$. In contrast, the interaction term, rewritten as
\begin{equation}
\label{eq:VdefCV}
\oV = \int \dd{\bf r}_a^3\dd{\bf r}_b^3 \frac{V_0}{8r_d^6(1-\beta\cos^2(\theta_d))^3} \frac{\os_{rr}({\bf r}_a)\os_{rr}({\bf r}_b)}{2}
\end{equation}
where $V_0=C_6/\sigma^6$ and  $\beta=1-\sigma_z^2/\sigma^2<1$ is the cloud's ellipticity, couples $\{\oP\}$ to many high-lying doubly-excited states. We treat these couplings as irreversible decays in a continuum, and extract the corresponding Lamb shifts and decay rates from an effective non-Hermitian potential $\oV_{e}(z)=V_0[z-(\oP\oG(z)\oP)^{-1}]$ calculated via the resolvent $\oG(z)=(z-\oV/V_0)^{-1}$, setting $z$ to match a characteristic energy $z_e$ in $V_0$ units \cite{Cohen1998.ch3}. As shown in Supp. Mat.~C, writing ${\bf k}=(n_c,l_c,n_d,l_d,m)$ for compactness,
\begin{multline}
\label{eq:ElemResolv}
\begin{array}{rcl}
\langle\Psi_{\bf k'}|\oG(z)|\Psi_{\bf k}\rangle 
&=& \delta_{n_c,n_c'}\delta_{l_c,l_c'}\delta_{m,m'} G_{n_d,l_d;n_d',l_d';m}(z),\\
z G_{n,l;n',l';m}(z) &=& \delta_{n,n'}\delta_{l,l'} + \end{array}
\\
(-1)^m\mathcal{N}_{n,2l}\mathcal{N}_{n',2l'}\sqrt{(4l+1)(4l'+1)}
\\
\sum_{l''=|l-l'|}^{l+l'}(4l''+1)
\left(\begin{array}{ccc}2l&2l'&2l''\\-m&m&0\end{array}\right)
\left(\begin{array}{ccc}2l&2l'&2l''\\0&0&0\end{array}\right)\\
\sum_{j=0}^n\left(\begin{array}{c}l-l'+n-j-1\\n-j\end{array}\right)
\sum_{j'=0}^{n'}\left(\begin{array}{c}l'-l+n'-j'-1\\n'-j'\end{array}\right)\\
\int_0^1\dd u P_{2l''}(u)Q_{j,j',l+l'}\left(\frac{z^{-1/3}}{4(1-\beta u^2)}\right),
\end{multline}
where $\mathcal{N}_{n,l}=\sqrt{2^n n!/(2n+2l+1)!!}$, $P_l$ is a Legendre polynomial, and the elements $Q_{j,j',l}(x) = Q_{j',j,l}(x)$ of $Q$, involving the Faddeeva function $ w(z)=\ee^{-z^2}\erfc(-iz)$, are obtained recursively:
\begin{align}
\nonumber
Q(x)
=&  ~\frac{1}{3}T(x)+\frac{2}{3}\Real\left(T\left(x\ee^{-2i\pi/3}\right)\right),\\
\nonumber
T_{0,0,0}(x)
=& ~
2x\left[1-i\sqrt{\pi x} w(-\sqrt{x})\right],\\
\nonumber
T_{0,0,l+1}(x)
=& ~
2x\left(T_{0,0,l}(x)+(2l+1)!!\right),\\
\nonumber
T_{j+1,j',l}(x)
=&
-\frac{x \mathcal{N}_{j,l}^{-2}}{j+1}\delta_{j,j'}
-\frac{j+l+1/2}{j+1}T_{j-1,j',l}(x)\\
& +
\frac{2j+l+3/2-x}{j+1}T_{j,j',l}(x).
\end{align}
In the spherical case, this simplifies as
\begin{multline}
\label{eq:ElemResolvSpher}
z\langle\Psi_{n_c',n_d',l'}|\oG(z)|\Psi_{n_c,n_d,l}\rangle  =
\delta_{n_c,n_c'}\delta_{l,l'}
\\
 \left(\delta_{n_d,n_d'}
+\mathcal{N}_{n_d,2l}\mathcal{N}_{n_d',2l}
Q_{n_d,n_d',2l}\left(z^{-1/3}/4\right)\right).
\end{multline}

To correctly reproduce the system's dynamics, we are left with the choice of the characteristic energy $z=z_e$.  Besides the driving strength $z_\Omega=\Omega/(2V_0)$, a relevant choice, providing a time-independent $\oV_e$, is the value $z_0$ that maximizes the density of doubly-excited states \cite{Bariani2012}
\begin{align}
\nonumber
p(z)=& \int\frac{\dd {\bf r}^3}{(2\pi)^{3/2}}\ee^{-r^2/2}\delta\left(\frac{1}{8r^6(1-\beta\cos^2\theta)^3}-z\right)\\
\label{eq:DOS}
=&~ \frac{\ee^{-1/(4z^{1/3})}}{12\sqrt{\beta}z^{4/3}} \erf\left(\sqrt{\frac{\beta}{1-\beta}}\frac{1}{2z^{1/6}}\right).
\end{align}
For a spherical cloud, $z_0=(3\sqrt{2})^{-6}\approx 1.7\times 10^{-4}$. It slightly increases when the cloud flattens, up to $4^{-6}\approx 2.4\times 10^{-4}$ when $\sigma_z/\sigma\rightarrow 0$. As $z_0 V_0= z_0 C_6/\sigma^6$ matches the typical interaction energies observed in experiments, we define the strong blockade condition in a Gaussian cloud as
$\Omega \ll \Omega_B=2 C_6/R_e^6$ with $R_e=3\sqrt{2}\sigma$. In this regime, the interactions largely exceed the power-broadened linewidth and prevent double Rydberg excitations. In contrast, using the radius $\sigma$ or the diameter $2\sigma$ as the effective cloud's size $R_e$ overestimates the interactions $C_6/R_e^6$ by orders of magnitude.

Unlike $\oV$, which unphysically diverges at short range where the van-der-Waals approximation fails, $\oV_e=\hat{\Lambda}-i\hat{\Gamma}$ remains finite. Its Hermitian part $\hat{\Lambda}$ includes Lamb shifts, while the Hermitian and non-negative $\hat{\Gamma}=\sum_{\bf k,k'}\Gamma_{\bf k;k'}|\Psi_{\bf k}\rangle\langle\Psi_{\bf k'}|$ describes decays. We can now construct a master equation for the system's density matrix $\orho$, restricted to $\{\oP\}$ and complemented with a state $\ket{C}$ accounting for the continuum:
\begin{align}
\nonumber
\dot{\orho} =& -i\left[\frac{\Omega}{2}(\oS+\oSd) + \hat{\Lambda},\orho \right] 
+\sum_{{\bf k},{\bf k}'} 2 \Gamma_{{\bf k};{\bf k}'} \oD_{\bf k,k'}[\orho], \\
\label{eq:Master}
\oD_{\bf k,k'}[\orho] =& \ketbra{C}{\Psi_{\bf k'}}\orho\ketbra{\Psi_{\bf k}}{C}-\{\ketbra{\Psi_{\bf k}}{\Psi_{\bf k'}},\orho\}/2
\end{align}
For a spherical cloud, the sum $\sum_{\bf k,k'}$ in the decay term can be reduced to $\sum_{n_c,l}$, see Supp. Mat.~C.3.

\begin{figure}
\centering
	\includegraphics[width=85mm]{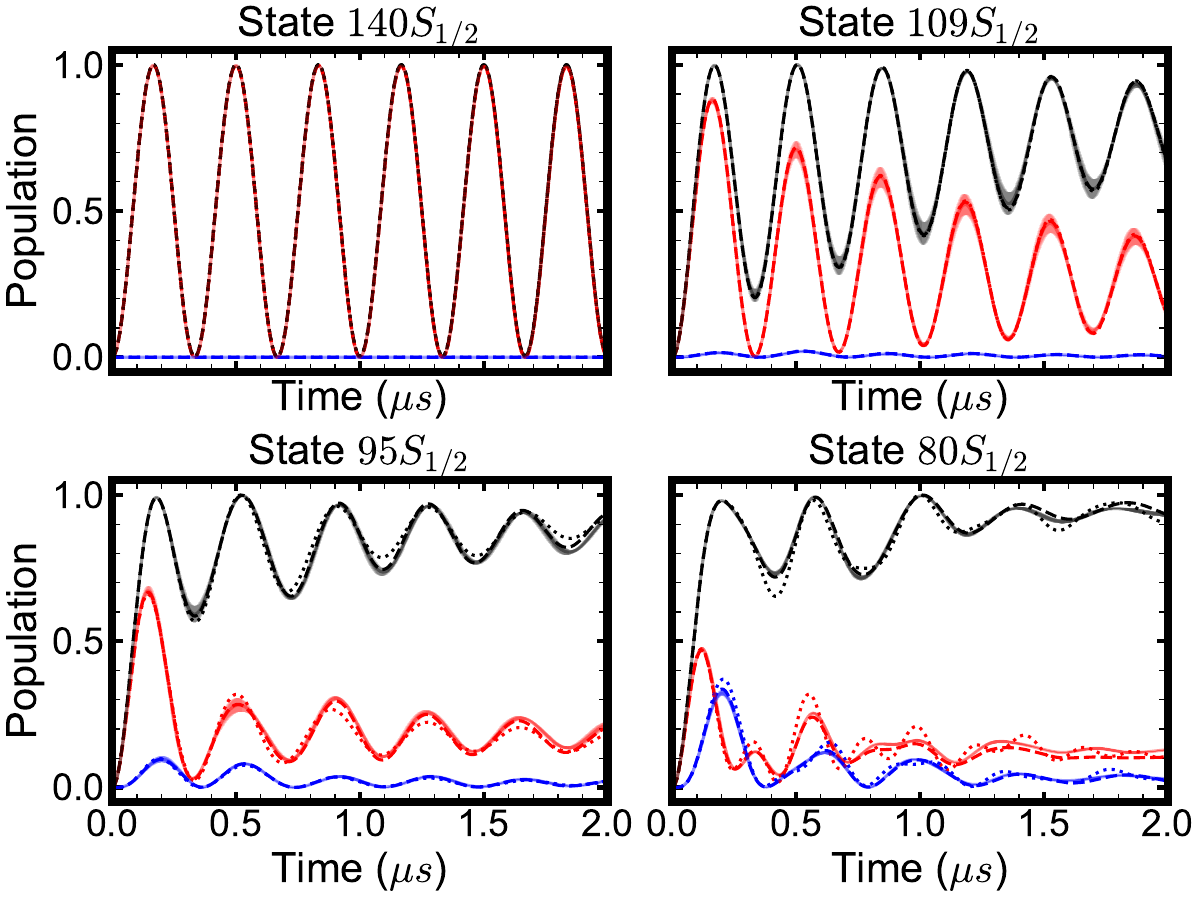}
	\caption{Theoretical simulations. Populations of the qubit's excited state $\ket{R}$ (red), of the doubly-excited symmetric state $\ket{\Psi_{0,0,0}}$ (blue), and of all excited states (black). A brute-force integration of Eq.~\ref{eq:HDriveDiscrete} (full lines) is compared with our model where $z_e=z_\Omega=\Omega/(2V_0)$ (dashed) or $z_e=z_0$ (dotted). The shaded areas around the full lines correspond to the standard deviations obtained from $8$ brute-force simulations. To speed up the latter, doubly-excited states with interaction energies exceeding $32\times\Omega/2$ are considered as far off-resonant and discarded from the Hilbert space.}
	\label{fig:CompBruteForce}
\end{figure}

\begin{figure*}
\centering
	\includegraphics[width=\textwidth]{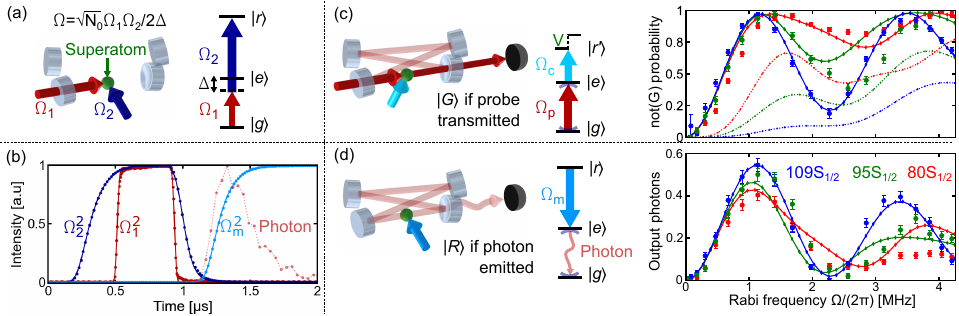}
	\caption{Experimental validation. (a) Rabi oscillations between the collective states \ket{G} and \ket{R} of a $^{87}$Rb superatomic qubit are driven via a two-photon transition coupling the atomic states $g=5S_{1/2}$ and $r=nS_{1/2}$ with n=(109,95,80). The collective Rabi frequency $\Omega=\sqrt{N_0}\Omega_1\Omega_2/2\Delta$ involves the effective number of atoms $N_0\approx 620$, the Rabi frequencies $\Omega_1\in\unit{[0,2\pi\times 22]}{\mega\hertz}$ and $\Omega_2=2\pi\times\unit{7.8}{\mega\hertz}$ on the two branches of the transition, and the detuning  $\Delta=-2\pi\times\unit{500}{\mega\hertz}$ from the intermediate state $e=5P_{1/2}$ (see text and Supp. Mat.~D). (b) Temporal shapes of the Rabi driving pulses, and those of the readout pulse and of the emitted photon used in the R/not(R) measurement shown in panel (d). (c) G/not(G) measurement. If, after the driving sequence, the superatom is in \ket{G}, a control beam coupling $e$ to the Rydberg state $r'=78S_{1/2}$ makes the cloud transparent to a probe resonant on the $g-e$ transition. Otherwise, Rydberg blockade shifts the $e-r'$ resonance and destroys the transparency. The contrast in the probe's transmission is enhanced by a resonant medium-finesse optical cavity. Circles show the not(G) probability, measured as a function of $\Omega$ for the states $109S_{1/2}$ (blue), $95S_{1/2}$ (green), and $80S_{1/2}$ (red). Error bars indicate the standard error. Solid lines represent the corresponding theoretical predictions, while dashed lines show the expected populations in the doubly-excited manifold. (d) R/not(R) measurement. Quasi-adiabatically increasing the Rabi frequency $\Omega_m$ of a mapping beam resonant on the $r-e$ transition coherently converts \ket{R} into a photon, emitted from the cavity in free space. The graph shows the measured number of emitted photons, corrected for detection losses outside the vacuum chamber, together with the theoretical predictions, with the same conventions as the plot above.}
	\label{fig:ExpData}
\end{figure*}

Fig.~\ref{fig:CompBruteForce} compares this model's predictions with brute-force simulations of Eq.~\ref{eq:HDriveDiscrete} using $N=400$ $^{87}$Rb atoms randomly distributed in a $\sigma=\sigma_z=\unit{5}{\micro\meter}$ Gaussian cloud and driven to Rydberg states 140S$_{1/2}$, 109S$_{1/2}$, 95S$_{1/2}$ and 80S$_{1/2}$, ranging from very strong to very weak blockade with $\Omega=2\pi\times\unit{3}{\mega\hertz}=(0.051,0.91,4.4,33)\Omega_B$ respectively \cite{Sibalic2017}. Considering two excitations at most and setting $n_{max}=(2,3,4,6)$ respectively reduces the Hilbert space from $\approx 80$k to $(12,19,29,59)$ states. The results are visually indistinguishable for 140S$_{1/2}$, 109S$_{1/2}$ and, when $z_e=z_\Omega$, even for the weakly blockaded state 95S$_{1/2}$ where the decay of the Rabi oscillations is fast and non-exponential, $z_e=z_0$ showing there a minor discrepancy.
A modest increase of the Hilbert space provides a good agreement even for the very weakly blockaded state 80S$_{1/2}$. The weaker the interactions, the faster the cloud becomes blockaded by asymmetric excitations, unresponsive to the laser driving. Doubly-excited populations remain small, making the simulations self-consistent and showing that the first effect of imperfect blockade is to allow Raman-like scattering into the singly-excited asymmetric manifold.

We test our model experimentally using a cold cloud of $N\approx 800$ $^{87}$Rb atoms with a Gaussian rms radius $\sigma = \unit{5.2}{ \micro \meter}$. We start by driving Rabi oscillations between \ket{G} and \ket{R} via a two-photon transition from the atomic ground state $g=5S_{1/2}$ to a Rydberg state $r=nS_{1/2}$ where $n\in \{80,95,109\}$, off-resonantly from the intermediate state $e=5P_{1/2}$ (see Fig.~\ref{fig:ExpData}(a) and Supp. Mat.~D). To minimize the effects of technical and thermal dephasing \cite{Covolo2025}, significant after a few $\mu$s, we drive the qubit for $\unit{500}{\nano\second}$ and control its rotation by changing the collective Rabi frequency $\Omega$ between $0$ and $2\pi\times\unit{4}{\mega\hertz}$. In comparison, $\Omega_B=2\pi\times\unit{\{0.072,0.538,2.64\}}{\mega\hertz}$ for the chosen states. Residual electric noise prevents us from reaching higher Rydberg states with stronger blockade. 

For each $n$ and $\Omega$, we then perform two binary measurements, G/not(G) and R/not(R). Their combination characterizes the superatom as a qubit \cite{Ebert2015,Spong2021,Xu2021,Yang2022,Vaneecloo2022}, a resource for two-photon quantum gates \cite{Stolz2022}, or a non-Gaussian light source \cite{Magro2023}. For G/not(G), we measure the transmission of a probe resonant on the $g-e$ transition in presence of a control beam coupling $e$ with $r'=78S_{1/2}$. If the superatom is in \ket{G}, the control laser makes it transparent. Otherwise, Rydberg blockade shifts the resonance of $r'$ and the transmission drops (see Fig.~\ref{fig:ExpData}(c), Supp. Mat.~D and Ref.~\cite{Vaneecloo2022}). In the R/not(R) measurement in Fig.~\ref{fig:ExpData}(d), we ramp up a mapping beam resonant on the $r-e$ transition and exploit the strong coupling between the cloud and a medium-finesse optical cavity to coherently convert \ket{R} into an optical photon emitted from the cavity into a well-defined free-propagating mode \cite{Magro2023}. As detailed in Supp. Mat.~E, in addition to imperfect blockade, the simulations of the experimental data in Fig.~\ref{fig:ExpData} account for laser noise, thermal dephasing \cite{Covolo2025} and, in the R/not(R) case, for the finite superatom-cavity coupling, the cavity extraction efficiency, the non-adiabatic mapping, and wavefront distortions of the driving beams, resulting in an overall $64\%$ mapping efficiency between \ket{R} and a photon propagating outside the vacuum system.

The experiments confirm the trends observed in Fig.~\ref{fig:CompBruteForce}.
When $\Omega$ increases, the blockade becomes weaker, allowing population to accumulate in asymmetric states: in Fig.~\ref{fig:ExpData}(c), the not(G) probability tends to $1$. These asymmetric excitations induce blockade and dephasing on the symmetric state, damping the Rabi oscillations between $\ket{G}$ and $\ket{R}$ in Fig.~\ref{fig:ExpData}(d). Our model closely matches the experimental data for the highest, moderately well blockaded 109S$_{1/2}$ Rydberg state, and reproduces its general features for weak blockades. Given the agreement with brute-force simulations in Fig.~\ref{fig:CompBruteForce}, discrepancies with the data can be attributed to effects beyond the model's assumptions. In particular, the van-der-Waals approximation neglects the complexity of Rydberg interactions at short range. In addition, for weak blockade Fig.~\ref{fig:ExpData}(c) predicts a significant population in the doubly-excited manifold, which makes triple excitations the main suspect.

We have shown, theoretically and experimentally, that the complex behavior of imperfectly-blockaded Rydberg superatoms can be reproduced by a numerically efficient model derived from first principles in a physically intuitive way. This ability is essential to adjust experimental parameters \cite{Vaneecloo2022}, optimize gates \cite{Mueller2009,Guo2020,Aggarwal2026}, or minimize leakage via dissipation engineering. The model is not specific to van-der-Waals interactions in Rydberg gases and can be generalized to ensembles of emitters interacting with a known, possibly anisotropic, potential. For example, resonant dipole-dipole interactions are successfully described in Supplemental Material C.4. It can also be transposed to multiple superatoms, e.g. to describe parity checks \cite{Guo2020}.

The model shows that single asymmetric excitations signal leakage errors more strongly than double 
ones. It also explains the relative weakness of Rydberg blockade in Gaussian superatoms: $R_e/\sigma=3\sqrt{2}\approx 4.2$ is of the order of $1$ as expected from dimensional analysis, but $C_6/R_e^6$ is weaker than $C_6/\sigma^6$ by nearly four orders of magnitude. Preliminary simulations confirm that removing the atoms located at the edges of the cloud strongly improves the blockade without significantly degrading the superatom-photon coupling. The generalization of this model to non-Gaussian density profiles, mitigating this issue, is underway.

\nocite{Gorini1976,Lindblad1976,qutip2}

\acknowledgments{
This work was funded by the ERC Starting Grant 677470 SEAQUEL, the Plan France 2030 project ANR-22-PETQ-0013, the Ile-de-France DIM SIRTEQ project CRIMP, and the CIFAR Azrieli Global Scholars program.}


%

\pagebreak
\onecolumngrid

\appendix

\section{3D harmonic oscillator eigenstates}
\label{app:SpherStates}

To describe spatial excitations in the atomic cloud, our model uses the eigenstates of a three-dimensional isotropic harmonic oscillator. To construct them algebraically, let us first introduce the one-dimensional bosonic creation operator $\oad_x = \ox/2 -i \op_x=x/2-\partial_x$ where $\op_x=-i\partial_x$ is the momentum along $x$, as well as $\oad_y$ and $\oad_z$ for the $y$ and $z$ coordinates. They satisfy $[\oa_i,\oad_{j}]=\delta_{i,j}$, and their conjugates annihilate the fundamental fully-symmetric state $\ket{\psi_{0,0,0}}$ with $\psi_{0,0,0}({\bf r})=\ee^{-{\bf r}^2/4}/(2\pi)^{3/4}$ . The ``Cartesian'' eigenstates of the Hamiltonian $\oH_{3D}=\oad_x\oa_x+\oad_y\oa_y+\oad_z\oa_z$ are thus of the form $(\oad_x)^{n_x}(\oad_y)^{n_y}(\oad_z)^{n_z}\ket{\psi_{0,0,0}}/\sqrt{n_x!n_y!,n_z!}$, with corresponding eigenvalues $n_x+n_y+n_z$. The structure of ``cylindrical'' eigenstates is the same if one replaces $\oad_x$ and $\oad_y$ with $\oad_\rho=(\oad_x+i\oad_y)/\sqrt{2}$ and  $\oad_{\rho^*}=(\oad_x-i\oad_y)/\sqrt{2}$. Expressing the projections $\oL_x$, $\oL_y$ and $\oL_z$ of the angular momentum $\hat{\bf L}$ as $\oL_i=\epsilon_{ijk}\hat{r}_{j}\op_k=-i\epsilon_{ijk}\oad_j\oa_k$, and observing that $\oL_z=\oad_\rho\oa_\rho-\oad_{\rho^*}\oa_{\rho^*}=\on_\rho-\on_{\rho^*}$ where $\on_j=\oad_j\oa_j$, shows that $\oad_\rho$ and $\oad_{\rho^*}$ create excitations with, respectively, $+1$ and $-1$ unit of angular momentum along $z$.

In spherical coordinates, the states  $\ket{\psi_{n,l,m}}$ with eigenfunctions 
\begin{align}
\label{eq:HOSpherFunc}
\psi_{n,l,m}({\bf r}) =&~ R_{n,l}(r)Y_l^m(\theta,\phi),& 
R_{n,l}(r) =&~ \sqrt[4]{\frac{2}{\pi}}\mathcal{N}_{n,l}r^l\ee^{-r^2/4} L_n^{(l+1/2)}\left(\frac{r^2}{2}\right),&
\mathcal{N}_{n,l} =&~ \sqrt{\frac{2^{n}n!}{(2n+2l+1)!!}},
\end{align}
conveniently separate into a radial part $R_{n,l}$ involving generalized Laguerre polynomials $L_n^{(l+1/2)}$, multiplied by a spherical harmonic $Y_l^m$. They co-diagonalize $\oH_{3D}$, $\oL^2$ and $\oL_z$ for the eigenvalues $2n+l,l(l+1)$ and $m$, respectively. To construct them algebraically, we define the usual angular momentum ladder operators $\oL_+=\oL_x+i\oL_y=\sqrt{2}(\oad_z\oa_{\rho^*}-\oad_\rho\oa_z)$ and $\oL_-=\oL_x-i\oL_y=\sqrt{2}(\oad_{\rho^*}\oa_z-\oad_z\oa_\rho)$ such that $\oL_\pm|\psi_{n,l,m}\rangle=\sqrt{l(l+1)-m(m\pm 1)}|\psi_{n,l,m\pm 1}\rangle$, and verify that $\oL_+(\oad_\orho)^l\ket{\psi_{0,0,0}}=0$ and $\oL_z(\oad_\orho)^l\ket{\psi_{0,0,0}}=l(\oad_\orho)^l\ket{\psi_{0,0,0}}$. With the Condon-Shortley convention, we thus have $\ket{\psi_{0,l,l}}=(-\oad_\rho)^l\ket{\psi_{0,0,0}}/\sqrt{l!}$. Writing $l(l+1)-m(m-1)=(l-m+1)(l+m)$, we get, for $|m|\leq l$,
\begin{equation}
\ket{\psi_{0,l,m}}
= \prod_{k=0}^{l-m-1}\frac{1}{\sqrt{(2l-k)(k+1)}}\oL_-^{l-m} \ket{\psi_{0,l,l}}
= \sqrt{\frac{(l+m)!}{l!(2l)!(l-m)!}}\oL_-^{l-m} (-\oad_\rho)^l \ket{\psi_{0,0,0}}.
\end{equation}
For the radial part, we introduce the operator 
$\oK=\oa_x^2+\oa_y^2+\oa_z^2=2\oa_\rho\oa_{\rho^*}+\oa_z^2$ which commutes with $\hat{\bf L}$, preserving $l$ and $m$. Because $\oH_{3D}=\oad_\rho\oa_\rho+\oad_{\rho^*}\oa_{\rho^*}+\oad_z\oa_z=\on_\rho+\on_{\rho^*}+\on_z$ verifies $[\oH_{3D},\oKd]=2\oKd$, we have $\oH_{3D}\oKd\ket{\psi_{n,l,m}}=(2n+l+2)\oKd\ket{\psi_{n,l,m}}$. Thus, as $\oKd$ preserves $l$, $\oKd\ket{\psi_{n,l,m}}\propto\ket{\psi_{n+1,l,m}}$. To determine the normalization factor, we expand $\oK\oKd 
=4\on_\rho\on_{\rho^*}+4\on_\rho+4\on_{\rho^*}+\on_z^2+3\on_z+2(\oad_z)^2\oa_\rho\oa_{\rho^*}+2\oad_\rho\oad_{\rho^*}(\oa_z)^2+6$
and use $\oL_+\oL_-=\oL_x^2+\oL_y^2-i[\oL_x,\oL_y]=\oL^2-\oL_z^2+\oL_z$, which gives
\begin{align}
\nonumber
\oL^2
=&~
2(\on_\rho\on_z+\on_{\rho^*}\on_z-\on_\rho\on_{\rho^*}) -2(\oad_z)^2\oa_\rho\oa_{\rho^*}-2\oad_\rho\oad_{\rho^*}(\oa_z)^2+\on_\rho^2+\on_{\rho^*}^2+\on_\rho+\on_{\rho^*}+2\on_z\\
\nonumber
\Rightarrow \oK\oKd + \oL^2  
=&~
\on_\rho^2+\on_{\rho^*}^2+\on_z^2+5(\on_\rho+\on_{\rho^*}+\on_z)+2\on_\rho\on_{\rho^*}+2\on_\rho\on_z+2\on_{\rho^*}\on_z+6 
=(\oH_{3D}+2)(\oH_{3D}+3)\\
\nonumber
\Rightarrow
\oK\oKd \ket{\psi_{n,l,m}}
=&~
[(2n+l+2)(2n+l+3)-l(l+1)]\ket{\psi_{n,l,m}}
= 2(n+1)(2(n+1)+2l+1)\ket{\psi_{n,l,m}}
\end{align}
The sign convention for Laguerre polynomials then gives $\ket{\psi_{n,l,m}} = -\oKd\ket{\psi_{n-1,l,m}}/\sqrt{2n(2n+2l+1)}$. Recursively,
\begin{align}
\nonumber
\ket{\psi_{n,l,m}}
=&~ 
\sqrt{\frac{(2l+1)!!}{2^{n}n!(2n+2l+1)!!}}(-\oKd)^{n}\ket{\psi_{0,l,m}}
= \mathcal{K}_{n,l,m}(-\oKd)^{n}\oL_-^{l-m}(-\oad_\rho)^l \ket{\psi_{0,0,0}},\\
\label{eq:HOSpherAlgDef}
\mathcal{K}_{n,l,m}
=&~ 
\sqrt{\frac{(l+m)!}{l!(2l)!(l-m)!}\frac{(2l+1)!!}{2^{n}n!(2n+2l+1)!!}}
=\sqrt{\left(\begin{array}{c}n+l\\n\end{array}\right)\frac{(l+m)!}{(l-m)!}\frac{2l+1}{l!(2n+2l+1)!}}.
\end{align}
In particular, we recover $\psi_{n,l,m}(-{\bf r}) = (-1)^l\psi_{n,l,m}({\bf r})$, as $\oKd$ and $\oL_-$ are invariant by parity while $\oad_\rho$ changes sign. 

Equation \ref{eq:HOSpherAlgDef} is the algeabraic definition of the 3D harmonic oscillator eigenstates we were looking for. It will be used in Supp. Mat.~\ref{app:DblExcDecomp} to calculate laser coupling matrix elements between singly- and doubly-excited states.

\section{Doubly-excited states in center-of-mass and relative coordinates}
\label{app:DblExcDecomp}

When the Rydberg blockade is imperfect, the laser can add a symmetric excitation to a singly-excited state $\ket{\psi_{n,l}}=\ket{\psi_{n,2l,0}}$, creating the doubly-excited state
\begin{equation}
\label{eq:SdagPsi}
\frac{\oSd}{\sqrt{1+\delta_{n+l}}}\ket{\psi_{n,l}}=\frac{1}{\sqrt{1+\delta_{n+l}}}\int \dd{\bf r}_a^3 \dd{\bf r}_b^3 \psi_{n,2l,0}({\bf r}_a)\psi_{0,0,0}({\bf r}_b)\os_{rg}({\bf r}_a)\os_{rg}({\bf r}_b)\ket{G}
\end{equation}
In this Supplemental Material, following the standard treatment of a two-body problem with interactions depending only on the relative position ${\bf r}_a-{\bf r}_b$, we decompose this state on states of the form
\begin{equation}
\label{eq:DblExcDiagAll}
\ket{\Phi_{n_c,l_c,m_c,n_d,l_d,m_d}}=\int \dd{\bf r}_a^3 \dd{\bf r}_b^3 \psi_{n_c,l_c,m_c}({\bf r}_c)\psi_{n_d,l_d,m_d}({\bf r}_d)\frac{\os_{rg}({\bf r}_a)\os_{rg}({\bf r}_b)}{\sqrt{2}}\ket{G}
\end{equation}
where, up to a factor $\sqrt{2}$, ${\bf r}_c=({\bf r}_a+{\bf r}_b)/\sqrt{2}$ is the position of the center of mass and ${\bf r}_d=({\bf r}_a-{\bf r}_b)/\sqrt{2}$ is the relative position of the excitations. One could expand the wavefunctions $\psi_{n,l,m}({\bf r})$ given by Eq.~\ref{eq:HOSpherFunc} analytically, but their operator-based definition in Eq.~\ref{eq:HOSpherAlgDef} significantly simplifies this calculation.

Let us leave aside Rydberg excitations for a moment, and consider two three-dimensional harmonic oscillators $a$ and $b$. The oscillator $a$ is in the state $|\psi_{n,2l,0}\rangle_a=\mathcal{K}_{n,2l,0} (-\oKd_a)^{n}\oL_{a-}^{2l}(-\oad_\rho)^{2l} |\psi_{0,0,0}\rangle_a$ given in Eq.~\ref{eq:HOSpherAlgDef}, where the operators
\begin{align}
\oad_\rho =&~ \frac{\oad_x+i\oad_y}{\sqrt{2}},&
\oad_{\rho^*} =&~ \frac{\oad_x-i\oad_y}{\sqrt{2}},&
\oL_{a-} =&~ \oLd_{a+}=\sqrt{2}(\oad_{\rho^*}\oa_z-\oad_z\oa_\rho),&
\oK_a =&~ 2\oa_\rho\oa_{\rho^*}+\oa_z^2
\end{align}
are defined via the 
bosonic 3D creation operator ${\bf \oad}=\hat{\bf r}_a/2-i\hat{\bf p}_a$ linear in position $\hat{\bf r}_a$ and momentum $\hat{\bf p}_a$. The oscillator $b$ is in the ground state $|\psi_{0,0,0}\rangle_b$, defined the same way via the 3D creation operator ${\bf \obd}$. From a quantum optics perspective, the coordinate rotation $(\hat{\bf r}_a, \hat{\bf r}_b) \rightarrow (\hat{\bf r}_c,\hat{\bf r}_d)$ corresponds to a beamsplitter transformation, changing ${\bf \oad}$ and ${\bf \obd}$ in ${\bf \ocd}=({\bf \oad}+{\bf \obd})/\sqrt{2}$ and ${\bf \odd}=({\bf \oad}-{\bf \obd})/\sqrt{2}$ which define the states $|\psi_{n_c,l_c,m_c}\rangle_c$ and $|\psi_{n_d,l_d,m_d}\rangle_d$.

Using $\psi_{n,l,m}(-{\bf r}) = (-1)^l\psi_{n,l,m}({\bf r})$, 
\begin{multline}
||\ket{\Phi_{n_c,l_c,m_c,n_d,l_d,m_d}}||^2
= \frac{1}{2}
\int \dd{\bf r}_a^3 \dd{\bf r}_b^3 \dd{{\bf r}'_a}^3 \dd{{\bf r}'_b}^3
\psi_{n_c,l_c,m_c}^*({\bf r}_c)\psi_{n_c,l_c,m_c}({\bf r}'_c)\psi_{n_d,l_d,m_d}^*({\bf r}_d)
\psi_{n_d,l_d,m_d}({\bf r}'_d)\\
\bra{G}\os_{gr}({\bf r}_a)\os_{gr}({\bf r}_b)\os_{rg}({\bf r}'_a)\os_{rg}({\bf r}'_b)\ket{G}\\
= \frac{1}{2}
\int \dd{\bf r}_a^3 \dd{\bf r}_b^3
\psi_{n_c,l_c,m_c}^*({\bf r}_c)\psi_{n_c,l_c,m_c}({\bf r}_c)\psi_{n_d,l_d,m_d}^*({\bf r}_d)
[\psi_{n_d,l_d,m_d}({\bf r}_d)+\psi_{n_d,l_d,m_d}(-{\bf r}_d)] = \frac{1+(-1)^{l_d}}{2},
\end{multline}
so $l_d$ must be even. The even parity of $|\psi_{n,2l,0}\rangle_a$ and of $|\psi_{0,0,0}\rangle_b$, invariant by ${\bf \oa}\leftrightarrow -{\bf \oa}$ and ${\bf \ob}\leftrightarrow -{\bf \ob}$, implies the same for  $|\psi_{n_c,l_c,m_c},\psi_{n_d,l_d,m_d}\rangle_{c,d}$ and thus for $l_c+l_d$; therefore $l_c$ and $l_d$ are both even and will be written as $2l_c$ and $2l_d$ from here on. It also leads to the exchange symmetry ${\bf \oc} \leftrightarrow {\bf \od}$, allowing us to assume $l_c\leq l_d$ to simplify the following calculation. The conservation of energy, understood here as the total number of vibrational quanta, implies $n_c+l_c+n_d+l_d=n+l$. A last set of symmetries stems from the conservation of total angular momentum ${\bf \oL}_T={\bf \oL}_a+{\bf \oL}_b={\bf \oL}_c+{\bf \oL}_d$: as ${\bf \oL}_b|\psi_{0,0,0}\rangle_b=0$, $|\psi_{n,2l,0},\psi_{0,0,0}\rangle_{a,b}$ is an eigenstate of $\oL_T^2$ and $\oL_{Tz}$, with azimuthal and magnetic quantum numbers $2l$ and $0$. This implies $m_d=-m_c=m \in [-2l_c,2l_c]$ and $l_d-l_c\leq l \leq l_c+l_d$. Moreover, the (real) scalar product of this state with  $|\psi_{n_c,2l_c,-m},\psi_{n_d,2l_d,m}\rangle_{c,d}$  must be proportional to the Clebsch coefficient $\langle 2l_c,-m;2l_d,m|2l,0\rangle$, or to the equivalent but more symmetric Wigner 3j symbol:
\begin{align}
\nonumber
\mathcal{A}_{n,l;l_c,n_d,l_d,m}
=&~ _{a,b}\langle\psi_{n,2l,0}\psi_{0,0,0}|\psi_{n_c,2l_c,-m},\psi_{n_d,2l_d,m}\rangle_{c,d}\\
=&~ \int \dd^3{\bf r}_a\dd^3{\bf r}_b \psi_{n,2l,0}({\bf r}_a)\psi_{0,0,0}({\bf r}_b) \psi_{n_c,2l_c,-m}({\bf r}_c)\psi_{n_d,2l_d,m}({\bf r}_d)\\
\label{eq:OpDefCpl12}
=&~ (-1)^{n_c+n_d-n} \mathcal{K}_{n,2l,0}\mathcal{K}_{n_c,2l_c,-m}\mathcal{K}_{n_d,2l_d,m} \langle \oa_\rho^{2l} \oL_{a+}^{2l}\oK_a^n (\oKd_c)^{n_c}\oL_{c-}^{2l_c+m}(\ocd_\rho)^{2l_c} (\oKd_d)^{n_d} \oL_{d-}^{2l_d-m}(\odd_\rho)^{2l_d}\rangle \\
\label{eq:DefCpl12W3j}
=&~ \mathcal{F}_{n,l;l_c,n_d,l_d}
\left(\begin{array}{ccc}
2l_c & 2l_d & 2l \\ -m & m & 0
\end{array}\right)
\end{align}
where, in Eq.~\ref{eq:OpDefCpl12}, the expectation value is taken in the ground state $\ket{0}=|\psi_{0,0,0},\psi_{0,0,0}\rangle_{a,b}=|\psi_{0,0,0},\psi_{0,0,0}\rangle_{c,d}$.

To calculate $\mathcal{F}$, let us set $m=-2l_c$ and write 
\begin{align}
\nonumber
\mathcal{A}_{n,l;l_c,n_d,l_d,-2l_c} =&~(-1)^{n+n_c+n_d} \mathcal{K}_{n,2l,0}\mathcal{K}_{n_c,2l_c,2l_c}\mathcal{K}_{n_d,2l_d,-2l_c}\mathcal{B}, \\
\mathcal{B} =&~ \left\langle \oa_\rho^{2l} \oL_{a+}^{2l}\oK_a^n (\oKd_c)^{n_c}(\oKd_d)^{n_d}(\ocd_\rho)^{2l_c}  \oL_{d-}^{2l_d+2l_c}(\odd_\rho)^{2l_d}\right\rangle.
\end{align}
Let us then expand
\begin{equation}
\nonumber
\oKd_c = (\oad_\rho+\obd_\rho)(\oad_{\rho^*}+\obd_{\rho^*})+\frac{(\oad_z+\obd_z)^2}{2}
= \frac{\oKd_a+\oKd_b}{2}+\oad_\rho\obd_{\rho^*}+\obd_\rho\oad_{\rho^*}+\oad_z\obd_z.
\end{equation}
As there are no $\ob$ operators on the left of $\oKd_c$ in $\mathcal{B}$, all terms involving $\obd$ will vanish. The same happens when expanding $\oKd_d$ and $\oc_\rho$, leaving
\begin{equation}
\mathcal{B} = \frac{\langle \oa_\rho^{2l}\oL_{a+}^{2l}\oK_a^n
(\oKd_a)^{n_c+n_d}(\oad_\rho)^{2l_c}\oL_{d-}^{2l_c+2l_d}(\odd_\rho)^{2l_d}\rangle}{2^{n_c+n_d+l_c}}
=
\frac{\partial_{\beta}^{2l_d+2l_c}
\langle \oa_\rho^{2l}\oL_{a+}^{2l}\oK_a^n
(\oKd_a)^{n_c+n_d}(\oad_\rho)^{2l_c}\ee^{\beta\oL_{d-}}(\odd_\rho)^{2l_d}\rangle |_{\beta=0}}{2^{n_c+n_d+l_c}} .
\end{equation}
As $[\oL_{d-},\odd_\rho]=-\sqrt{2}\odd_z$, $[\oL_{d-},[\oL_{d-},\odd_\rho]]=-2\odd_{\rho^*}$, $[\oL_{d-},[\oL_{d-},[\oL_{d-},\odd_\rho]]]=0$ and $\oL_{d-}\ket{0}=0$, we can use the Hadamard formula
\begin{equation}
\ee^{\oA}\oB\ee^{-\oA} = \ee^{[\oA,\cdot]}\oB = \sum_{n=0}^{+\infty} \frac{1}{n!} [\oA,[\oA,...[\oA,\oB]]] = \oB+[\oA,\oB]+\frac{1}{2}[\oA,[\oA,\oB]]+...
\end{equation}
to commute the last two operators, remove all the $\obd$ like previously, then commute them back:
\begin{align}
\nonumber
\ee^{\beta\oL_{d-}}(\odd_\rho)^{2l_d}\ket{0}
 =&~ \left(\ee^{\beta\oL_{d-}}\odd_\rho\ee^{-\beta\oL_{d-}}\right)^{2l_d}\ket{0}
 = \left(\odd_\rho-\sqrt{2}\beta\odd_z-\beta^2\odd_{\rho^*}\right)^{2l_d}\ket{0}\\
\nonumber
 \Rightarrow
 \mathcal{B}
 =&~
\frac{1}{2^{n_c+n_d+l_c}} \partial_{\beta}^{2l_d+2l_c} \left.\left\langle \oa_\rho^{2l}\oL_{a+}^{2l}\oK_a^n
(\oKd_a)^{n_c+n_d}(\oad_\rho)^{2l_c}\left(\odd_\rho-\sqrt{2}\beta\odd_z-\beta^2\odd_{\rho^*}\right)^{2l_d}\right\rangle \right|_{\beta=0}\\
\nonumber
=&~
\frac{1}{2^{n_c+n_d+l_c+l_d}} \partial_{\beta}^{2l_d+2l_c} \left.\left\langle \oa_\rho^{2l}\oL_{a+}^{2l}\oK_a^n
(\oKd_a)^{n_c+n_d}(\oad_\rho)^{2l_c}\left(\oad_\rho-\sqrt{2}\beta\oad_z-\beta^2\oad_{\rho^*}\right)^{2l_d}\right\rangle \right|_{\beta=0}\\
=&~
\frac{1}{2^{n_c+n_d+l_c+l_d}}
\langle \oa_\rho^{2l}\oL_{a+}^{2l}\oK_a^n
(\oKd_a)^{n_c+n_d}(\oad_\rho)^{2l_c}\oL_{a-}^{2l_d+2l_c}(\oad_\rho)^{2l_d}\rangle.
\end{align}
From here on, we can drop the index $a$ in the operators. Verifying that 
\begin{equation}
\label{oKoLplusPsiNLM}
\oK^{p}\oL_+^q(\oKd)^n\oL_-^{l-m}(\oad_\rho)^l\ket{0} = \frac{\mathcal{K}_{n-p,l,m+q}^2}{\mathcal{K}_{n,l,m}^2}(\oKd)^{n-p}\oL_-^{l-m-q}(\oad_\rho)^l\ket{0}
\end{equation}
 by taking the scalar product with $(\oKd)^{n-p}\oL_-^{l-m-q}(\oad_\rho)^l\ket{0}$, and using $l_c+l_d+n_c+n_d=n+l$,
\begin{equation}
\mathcal{B}
= \frac{1}{2^{n+l}}\frac{\mathcal{K}_{l_c+l_d-l,2l,0}^2}{\mathcal{K}_{n,2l,0}^2} \left\langle \oa_\rho^{2l}\oL_{+}^{2l}\oK^{l_c+l_d-l}
(\oad_\rho)^{2l_c} \oL_-^{2l_c+2l_d} (\oad_\rho)^{2l_d}\right\rangle
\end{equation}
The operator $\oL_{+}$ commutes with $\oK$, but also with $\oad_\rho$. Reusing Eq.~\ref{oKoLplusPsiNLM} on $\oL_{+}^{2l}\oL_-^{2l_c+2l_d} (\oad_\rho)^{2l_d}\ket{0}$,
\begin{align}
\mathcal{B} 
=&~\frac{1}{2^{n+l}}
\frac{\mathcal{K}_{l_c+l_d-l,2l,0}^2}{\mathcal{K}_{n,2l,0}^2}
\frac{\mathcal{K}_{0,2l_d,2l-2l_c}^2}{\mathcal{K}_{0,2l_d,-2l_c}^2}\mathcal{C},
&
\mathcal{C} =&
\left\langle \oa_\rho^{2l}\oK^{l_c+l_d-l}
(\oad_\rho)^{2l_c} \oL_-^{2(l_c+l_d-l)} (\oad_\rho)^{2l_d}\right\rangle
\end{align}
Let us now use the Hadamard formula with $[\alpha\oa_\rho+\beta\oK,\oad_\orho]=\alpha+2\beta\oa_{\rho^*}$ and $[\alpha\oa_\rho+\beta\oK,[\alpha\oa_\rho+\beta\oK,\oad_\orho]]=0$:
\begin{align}
\nonumber
\mathcal{C} =&
\left.\partial_{\alpha}^{2l}\partial_\beta^{l_c+l_d-l}\left\langle \ee^{\alpha\oa_\rho+\beta\oK}
(\oad_\rho)^{2l_c} \oL_-^{2(l_c+l_d-l)} (\oad_\rho)^{2l_d}\right\rangle\right|_{\alpha=\beta=0}\\
=&
\left.\partial_{\alpha}^{2l}\partial_\beta^{l_c+l_d-l}\left\langle 
(\oad_\rho+\alpha+2\beta\oa_{\rho^*})^{2l_c} \ee^{\alpha\oa_\rho}\ee^{\beta\oK} \oL_-^{2(l_c+l_d-l)} (\oad_\rho)^{2l_d}\right\rangle\right|_{\alpha=\beta=0}.
\end{align}
As $\oK\oL_-^{l-m} (\oad_\rho)^{l}\ket{0}=0$, we can eliminate $\ee^{\beta\oK}$, as well as $\oad_\rho$ from the leftmost operator, leaving
\begin{align}
\nonumber
\mathcal{C}
=&~ \frac{(2l_c)!}{(l_c-l_d+l)!} 2^{l_c+l_d-l} 
\left.\partial_{\alpha}^{2l} \left\langle 
\oa_{\rho^*}^{l_c+l_d-l} \alpha^{l_c-l_d+l} \ee^{\alpha\oa_\rho} \oL_-^{2(l_c+l_d-l)} (\oad_\rho)^{2l_d}\right\rangle \right|_{\alpha=0}\\
=&~ \left(\begin{array}{c} 2l \\ l_c-l_d+l\end{array}\right) (2l_c)! 2^{l_c+l_d-l} \left\langle 
\oa_{\rho^*}^{l_c+l_d-l} \oa_\rho^{l+l_d-l_c} \oL_-^{2(l_c+l_d-l)} (\oad_\rho)^{2l_d}\right\rangle
\end{align}
A last use of the Hadamard formula on the two rightmost operators finally gives
\begin{align}
\mathcal{C}
\nonumber
=&~ \left(\begin{array}{c} 2l \\ l_c-l_d+l\end{array}\right) (2l_c)! 2^{l_c+l_d-l} \left.\partial_{\alpha}^{2(l_c+l_d-l)}  \left\langle 
\oa_{\rho^*}^{l_c+l_d-l} \oa_\rho^{l+l_d-l_c} \ee^{\alpha\oL_-} (\oad_\rho)^{2l_d}\right\rangle \right|_{\alpha=0}\\
\nonumber
=&~ \frac{(2l)!(2l_c)! 2^{l_c+l_d-l}}{(l+l_c-l_d)!(l+l_d-l_c)!} \left.\partial_{\alpha}^{2(l_c+l_d-l)}  \left\langle 
\oa_{\rho^*}^{l_c+l_d-l} \oa_\rho^{l+l_d-l_c} (\oad_\rho-\sqrt{2}\alpha\oad_z-\alpha^2\oad_{\rho^*})^{2l_d}\right\rangle \right|_{\alpha=0}\\
\nonumber
=&~ \frac{ (-1)^{l_c+l_d-l}(2l)!(2l_c)! 2^{l_c+l_d-l}}{(l+l_c-l_d)!(l+l_d-l_c)!}  \left(\begin{array}{c} 2l_d \\ l_c+l_d-l\end{array}\right)(l_c+l_d-l)!(l+l_d-l_c)!(2l_c+2l_d-2l)!, \\
\mathcal{B} =&~
\frac{(-1)^{l_c+l_d-l}}{2^{n+2l-l_c-l_d}} 
\frac{\mathcal{K}_{l_c+l_d-l,2l,0}^2}{\mathcal{K}_{n,2l,0}^2}
\frac{\mathcal{K}_{0,2l_d,2l-2l_c}^2}{\mathcal{K}_{0,2l_d,-2l_c}^2}
\frac{(2l_c)!(2l_d)!(2l)!(2l_c+2l_d-2l)!}{(l_c-l_d+l)!(l_d-l_c+l)!}.
\end{align}
Reverting from $\mathcal{B}$ to $\mathcal{A}$, we simplify $(-1)^{n_c+n_d-n}(-1)^{l_c+l_d-l}=1$ and get 
\begin{equation}
\label{EqACircInterm}
\mathcal{A}_{n,l;l_c,n_d,l_d,-2l_c} 
= \frac{1}{2^{n+2l-l_c-l_d}}
\frac{\mathcal{K}_{n_c,2l_c,2l_c}\mathcal{K}_{n_d,2l_d,-2l_c}}{\mathcal{K}_{n,2l,0}}
\frac{\mathcal{K}_{l_c+l_d-l,2l,0}^2\mathcal{K}_{0,2l_d,2l-2l_c}^2}{\mathcal{K}_{0,2l_d,-2l_c}^2}
\frac{(2l_c)!(2l_d)!(2l)!(2l_c+2l_d-2l)!}{(l_c-l_d+l)!(l_d-l_c+l)!}.
\end{equation}
The Racah formula gives the $m$-dependent Wigner 3j symbol that we need to factor out:
\begin{equation}
\left(\begin{array}{ccc}
2l_c & 2l_d & 2l \\ 2l_c & -2l_c & 0
\end{array}\right)
 = \sqrt{\frac{(4l_c)!(2l_d+2l_c)!(2l+2l_d-2l_c)!}{(2l+2l_d+2l_c+1)!(2l+2l_c-2l_d)!(2l_c+2l_d-2l)!(2l_d-2l_c)!}}.
\end{equation}
As we added angular momenta, we can also expect to see appearing
\begin{equation}
\left(\begin{array}{ccc}
2l_c & 2l_d & 2l \\ 0 & 0 & 0
\end{array}\right)
 = \sqrt{\frac{(2l_d+2l_c-2l)!(2l+2l_c-2l_d)!(2l+2l_d-2l_c)!}{(2l+2l_d+2l_c+1)!}}
 \frac{(-1)^{l_c+l_d-l}(l+l_d+l_c)!}{(l_c+l_d-l)!(l+l_c-l_d)!(l+l_d-l_c)!}.
\end{equation}
Using Eq.\ref{eq:HOSpherAlgDef} to expand all the $\mathcal{K}_{n,l,m}$ in Eq. \ref{EqACircInterm}, this allows one to group many factors and get
\begin{align}
\nonumber
\mathcal{A}_{n,l;l_c,n_d,l_d,-2l_c} 
&= (-1)^{l_c+l_d-l}\frac{4l+1}{2^{n+l}}\sqrt{\frac{\mathcal{D}_{n_c,l_c}\mathcal{D}_{n_d,l_d}}{\mathcal{D}_{n,l}}}
\left(\begin{array}{ccc}
2l_c & 2l_d & 2l \\ 0 & 0 & 0
\end{array}\right)
\left(\begin{array}{ccc}
2l_c & 2l_d & 2l \\ 2l_c & -2l_c & 0
\end{array}\right)\\
\nonumber
\Rightarrow 
\mathcal{A}_{n,l;l_c,n_d,l_d,m} 
&= (-1)^{l_c+l_d-l}\frac{4l+1}{2^{n+l}}\sqrt{\frac{\mathcal{D}_{n_c,l_c}\mathcal{D}_{n_d,l_d}}{\mathcal{D}_{n,l}}}
\left(\begin{array}{ccc}
2l_c & 2l_d & 2l \\ 0 & 0 & 0
\end{array}\right)
\left(\begin{array}{ccc}
2l_c & 2l_d & 2l \\ -m & m & 0
\end{array}\right),\\
\mathcal{D}_{n,l} &= \frac{4l+1}{2^n n!(2n+4l+1)!!} =  \frac{(4l+1)2^{2l}(n+2l)!}{n!(2n+4l+1)!},
\end{align}
where energy conservation $n_c+n_d+l_c+l_d=n+l$ is still assumed but $l_c\leq l_d$ can be relaxed.

Returning to doubly-excited Rydberg states,
\begin{equation}
\label{eq:DblExcExpansUngrp}
\oSd\ket{\psi_{n,l}} = \sqrt{2}\sum_{l_c,n_d,l_d,m}\mathcal{A}_{n,l;l_c,n_d,l_d,m} \ket{\Phi_{n_c,2l_c,-m,n_d,2l_d,m}},
\end{equation}
where the sum runs over $n_c+n_d+l_c+l_d=n+l$, $|l_c-l_d|\leq l \leq l_c+l_d$ and $|m|\leq \min(2l_c,2l_d)$. As the interaction potential is invariant by rotation around $Oz$ and thus block-diagonal in $m$, and as
\begin{equation}
\label{eq:W3jMinusM}
\left(\begin{array}{ccc}
l_1 & l_2 & l_3 \\ -m_1 & -m_2 & -m_3
\end{array}\right)
=(-1)^{l_1+l_2+l_3}
\left(\begin{array}{ccc}
l_1 & l_2 & l_3 \\ m_1 & m_2 & m_3
\end{array}\right)
\end{equation}
implies $\mathcal{A}_{n,l;l_c,n_d,l_d,m}=\mathcal{A}_{n,l;l_c,n_d,l_d,-m}$, we can group these terms two by two to further reduce the dimension of the doubly-excited subspace, and restrict the sum over $m$ to $0\leq m \leq \min(2l_c,2l_d)$ to get
\begin{align}
\oSd\ket{\psi_{n,l}}
=&~ \sum_{n_d,l_c,l_d,m\geq 0} S_{n,l;l_c,n_d,l_d,m} \ket{\Psi_{n_c,l_c,n_d,l_d,m}},\\
S_{n,l;l_c,n_d,l_d,m}
=&~
\frac{2}{\sqrt{1+\delta_{m}}} \mathcal{A}_{n,l;l_c,n_d,l_d,m},\\
\ket{\Psi_{n_c,l_c,n_d,l_d,m}}
=&~ \frac{\ket{\Phi_{n_c,2l_c,-m,n_d,2l_d,m}}+\ket{\Phi_{n_c,2l_c,m,n_d,2l_d,-m}}}{\sqrt{2(1+\delta_m)}},
\end{align}
giving Equations 3 and 4 in the main text.

Since the transformation between $({\bf \oad},{\bf \obd})$ and $({\bf \ocd},{\bf \odd})$ is unitary, $\ket{\Psi_{n_c,l_c,n_d,l_d,m}}$ can be decomposed on states with wavefunctions of the form $\psi_{n_a,l_a,m_a}({\bf r}_a)\psi_{n_b,l_b,m_b}({\bf r}_b)$, annihilated by $\oS$ unless $n_a=l_a=m_a=0$ or $n_b=l_b=m_b=0$ (without loss of generality, we will assume the former). Then, energy and angular momentum conservation show that $m_b=0$, $l_b=2l$ with $|l_c-l_d|\leq l \leq l_c+l_d$, and $n_b=n$ with $n+l=n_c+n_d+l_c+l_d$. Thus
\begin{equation}
\oS\ket{\Psi_{n_c,l_c,n_d,l_d,m}}
= \sum_{n,l} S_{n,l;l_c,n_d,l_d,m} \ket{\psi_{n,l}}
= \sum_{l=|l_c-l_d|}^{l_c+l_d} S_{n_c+n_d+l_c+l_d-l,l;l_c,n_d,l_d,m} \ket{\psi_{n_c+n_d+l_c+l_d-l,l}}
\end{equation}
and, in the limit of two Rydberg excitations, the Hilbert space formed by $\ket{G}$, $\{\ket{\psi_{n,l}}\}$ and $\{\ket{\Psi_{n_c,l_c,n_d,l_d,m}}\}$ with $n+l\leq n_{max}$ and $n_c+n_d+l_c+l_d \leq n_{max}$ is stable under the action of $\oS$ and $\oSd$.

\bigskip
For a spherically-symmetric cloud, the singly-excited subspace is restricted to states $|\psi_{n}\rangle$ with zero angular momentum, which restricts the doubly-excited subspace to $l_c=l_d$, simplified as $l$ in the following. The rotational invariance of the interactions and
\begin{equation}
\mathcal{A}_{n,0;l,n_d,l,m} = 
\mathcal{F}_{n,0;l,n_d,l}\left(\begin{array}{ccc}
2l & 2l & 0 \\ -m & m & 0
\end{array}\right)
=\mathcal{F}_{n,0;l,n_d,l}\frac{(-1)^{m}}{\sqrt{4l+1}}
\end{equation}
allow us to reduce the dimension of the doubly-excited subspace by separating the sum over $m$ in Eq.~\ref{eq:DblExcExpansUngrp} to combine $4l+1$ states $\ket{\Psi_{n_c,l,n_d,l,m}}$ in a single state
\begin{equation}
\label{eq:StateDblExcSpher}
\ket{\Psi_{n_c,n_d,l}}
= \sum_{m=-2l}^{2l}\frac{(-1)^m}{\sqrt{4l+1}}\ket{\Phi_{n_c,l,-m,n_d,l,m}}
= \sum_{m=0}^{2l} \sqrt{\frac{2}{1+\delta_m}} \frac{(-1)^m}{\sqrt{4l+1}}\ket{\Psi_{n_c,l,n_d,l,m}},
\end{equation}
laser-coupled to $\ket{\psi_{n}}$ with $n=n_c+n_d+2l$:
\begin{align}
\oSd\ket{\psi_{n}}
=&~ \sum_{n_d,l} S_{n;n_d,l}\ket{\Psi_{n_c,n_d,l}},\\
\oS\ket{\Psi_{n_c,n_d,l}} =&~ S_{n;n_d,l} \ket{\psi_{n}}\\
S_{n;n_d,l}
=&~
\sqrt{2}\mathcal{F}_{n,0;l,n_d,l} = \frac{\sqrt{2}}{2^n}\sqrt{\frac{\mathcal{D}_{n_c,l}\mathcal{D}_{n_d,l}}{(4l+1)\mathcal{D}_{n,0}}}.
\end{align}
We thus recover Eq.~5 in the main text.

\section{Effective description of interactions in a restricted subspace}
\label{app:EffInteract}

In this Supplemental Material, we derive an effective non-Hermitian potential describing the effect of Rydberg interactions in a subspace. The decay rates obtained from its anti-Hermitian part are then injected in a standard master equation.

\subsection{Effective non-Hermitian Hamiltonian}

We restrict the Hilbert space of the system to a small subspace $\{\oP\}$ formed by the ground state $\ket{G}$, a few low-lying singly-excited states $\ket{\psi_{n,l}}$ with $n+l\leq n_{max}$, and a few low-lying doubly-excited states $\ket{\Psi_{n_c,l_c,n_d,l_d,m}}$ defined in Supp. Mat.~\ref{app:DblExcDecomp}, where all indices are positive integers, $0\leq m \leq \min(2l_c,2l_d)$ and $n_c+n_d+l_c+l_d\leq n_{max}$. This subspace corresponds to a projector $\oP=\ketbra{G}{G}+\sum_{n,l}\ketbra{\psi_{n,l}}{\psi_{n,l}}+\sum_{\bf k}\ketbra{\Psi_{\bf k}}{\Psi_{\bf k}}$ where we adopted the compact notation ${\bf k}=(n_c,l_c,n_d,l_d,m)$. The Rydberg interaction potential (Eq.~6 in the main text)
\begin{align}
\oV = \int \dd{\bf r}_a^3\dd{\bf r}_b^3 \frac{V_0}{8r_d^6(1-\beta\cos^2(\theta_d))^3} \frac{\os_{rr}({\bf r}_a)\os_{rr}({\bf r}_b)}{2},
\end{align}
couples these doubly-excited states to many higher-lying ones, treated here as a broad memory-less continuum. The coupling to this continuum leads to Lamb shifts and decays, which we obtain from an effective non-Hermitian potential $\oV_e=\hat{\Lambda}-i\hat{\Gamma}$.  Its Hermitian part $\hat{\Lambda}$ describes the unitary part of the interactions, including Lamb shifts, while $\hat{\Gamma}$ contains time-independent decay rates which we can integrate is a standard Gorini - Kossakowski - Sudarshan - Lindblad (GKSL) master equation \cite{Gorini1976,Lindblad1976} given in Eq.~11 of the main text.

To calculate $\oV_e$, we follow Ref.~\cite{Cohen1998.ch3} and use the resolvent of $\oV/V_0$:
\begin{align}
\oG(z) =&~\frac{1}{z-\oV/V_0}
= \int \dd{\bf r}_a^3\dd{\bf r}_b^3 G(z,{\bf r}_d) \frac{\os_{rr}({\bf r}_a)\os_{rr}({\bf r}_b)}{2},&
G(z,{\bf r}_d) =&~\frac{1}{z-1/[8r_d^6(1-\beta\cos^2(\theta_d))^{3}]}
\end{align}
In the limit $\Imag(z) \rightarrow 0^+$ assumed in the following, $\oG(z)$ is related to the forward propagator $\oU(t\geq 0)=\ee^{-i\oV t}$ via
\begin{align}
\label{eq:PropVsResolv}
\oG(z)=&-iV_0\int_0^{+\infty}\dd t\,\ee^{i (V_0z-\oV)t} = -iV_0\int_0^{+\infty}\dd t\, \oU(t)\ee^{i V_0zt},&
\oU(t) =&
 \frac{i}{2\pi}\int_{-\infty}^{+\infty}\dd z\, \oG(z)\ee^{-iV_0zt}.
\end{align}
Defining $\oV_e(z)= V_0[ z -(\oP\oG(z)\oP)^{-1}]$ gives the exact evolution in the subspace of interest as
\begin{equation}
\label{eq:PropRestrict}
\oP\oU(t)\oP = \frac{i}{2\pi}\int_{-\infty}^{+\infty}\dd z\, \frac{\oP}{z -\oV_e(z)/V_0}\ee^{-iV_0zt}
\end{equation}
In this integral, we then approximate $\oV_e(z)\approx \oV_e(z_{e})$ where $z_e$ is a characteristic energy rescaled by $V_0$. In our case, we can assume $\Real(z_e)>0$. The result of the integration then takes the desired form $\oP\oU(t)\oP = \ee^{-i\oV_e(z_e)t}$.

This reduces the problem to calculating the matrix elements of $\oG(z)$, choosing an appropriate $z_e$, and obtaining $\oV_e(z_e)$ by numerical inversion. Using $\psi_{n,l,m}(-{\bf r})=(-1)^l\psi_{n,l,m}({\bf r})$,
\begin{align}
\nonumber
\bra{\Psi_{\bf k'}}\oG(z)\ket{\Psi_{\bf k}}
=&~
\int\frac{\dd{\bf r}_a^3\dd{\bf r}_b^3\dd{\bf r}_a'^3\dd{\bf r}_b'^3\dd{\bf r}_a''^3\dd{\bf r}_b''^3}{8\sqrt{1+\delta_{m}}\sqrt{1+\delta_{m'}}}
[
\psi_{n_c',2l_c',-m'}^*({\bf r}_c')\psi_{n_d',2l_d',m'}^*({\bf r}_d')+\psi_{n_c',2l_c',m'}^*({\bf r}_c')\psi_{n_d',2l_d',-m'}^*({\bf r}_d')]\\
&\nonumber
G(z,{\bf r}_d'')[\psi_{n_c,2l_c,-m}({\bf r}_c)\psi_{n_d,2l_d,m}({\bf r}_d)+\psi_{n_c,2l_c,m}({\bf r}_c)\psi_{n_d,2l_d,-m}({\bf r}_d)]\\
&\nonumber
\bra{G}\os_{gr}({\bf r}_a')\os_{gr}({\bf r}_b')\os_{rr}({\bf r}_a'')\os_{rr}({\bf r}_b'')\os_{rg}({\bf r}_a)\os_{rg}({\bf r}_b)\ket{G}\\
\nonumber
=&\delta_{n_c,n_c'}\delta_{l_c,l_c'}\delta_{m,m'}
G_{n_d,l_d;n_d',l_d';m}(z),\\
\label{eq:ResolIntegr3D}
z G_{n,l;n',l';m}(z)
=&~z \int\dd{\bf r}^3\frac{\psi_{n',2l',m}^*({\bf r})\psi_{n,2l,m}({\bf r})}{z-1/[8 r^6(1-\beta\cos^2(\theta_d))^{3}]}
= \delta_{n,n'}\delta_{l,l'}+
\int\dd^3{\bf r}
\frac{\psi_{n',2l',m}^*({\bf r})\psi_{n,2l,m}({\bf r})}{8z r^6(1-\beta\cos^2(\theta_d))^{3}-1}.
\end{align}
This expression can be reduced to an integral of a non-divergent function of a single variable, easy to calculate numerically. For this, we can use Eq.~\ref{eq:HOSpherFunc} to write
$\psi_{n',2l',m}^*({\bf r})\psi_{n,2l,m}({\bf r})
= R_{n,2l}(r)R_{n',2l'}(r)  Y_{2l'}^{m*}(\theta,\phi)Y_{2l}^{m}(\theta,\phi)$, and apply the contraction rule of spherical harmonics
\begin{equation}
Y_{l_1}^{m_1}(\theta,\phi)Y_{l_2}^{m_2}(\theta,\phi)
=
\sqrt{\frac{(2l_1+1)(2l_2+1)}{4\pi}}\sum_{l=|l_1-l_2|}^{l_1+l_2}\sqrt{2l+1}
\left(\begin{array}{ccc}l_1&l_2&l\\m_1&m_2&-m_1-m_2\end{array}\right)
\left(\begin{array}{ccc}l_1&l_2&l\\0&0&0\end{array}\right)Y_{l}^{m_1+m_2}(\theta,\phi)
\end{equation}
to $Y_{2l'}^{m*}(\theta,\phi)Y_{2l}^{m}(\theta,\phi)=(-1)^m Y_{2l'}^{-m}(\theta,\phi)Y_{2l}^{m}(\theta,\phi)$. In the second Wigner 3j symbol, $l$ must then be even, giving
\begin{align}
\int_0^{2\pi}\dd\phi Y_{2l'}^{m*}(\theta,\phi)Y_{2l}^{m}(\theta,\phi)
=&
(-1)^m\sqrt{(4l+1)(4l'+1)}\sum_{l''=|l-l'|}^{l+l'}\frac{4l''+1}{2}
\left(\begin{array}{ccc}2l&2l'&2l''\\-m&m&0\end{array}\right)
\left(\begin{array}{ccc}2l&2l'&2l''\\0&0&0\end{array}\right)
P_{2l''}(\cos\theta)
\end{align}
where $P_{2l}(x)$ is a Legendre polynomial. In the radial part $ R_{n,2l}(r)R_{n',2l'}(r)$, the Laguerre polynomials can be converted to the same orthogonal family $L_{j}^{(l+l'+1/2)}$ by using
\begin{equation}
\label{EqLagNAlphaLagIBeta}
L_n^{(\alpha)}(x)=\sum_{j=0}^n\left(\begin{array}{c}\alpha-\beta+n-j-1\\n-j\end{array}\right)L_j^{(\beta)}(x)
\end{equation}
where the definition of the binomial coefficient
\begin{equation}
\left(\begin{array}{c}\alpha\\k\end{array}\right)=\frac{\alpha(\alpha-1)...(\alpha-k+1)}{k(k-1)...1}
\end{equation}
is valid for $k\in \mathbb{N}$ and $\alpha\in \mathbb{R}$. Replacing $u=\cos\theta$ in the spatial integral, we obtain
\begin{align}
\nonumber
z G_{n,l;n',l';m}(z)
= &
 \delta_{n,n'}\delta_{l,l'}
+(-1)^m\mathcal{N}_{n,2l}\mathcal{N}_{n',2l'}\sqrt{(4l+1)(4l'+1)}\sum_{l''=|l-l'|}^{l+l'}(4l''+1)
\left(\begin{array}{ccc}2l&2l'&2l''\\-m&m&0\end{array}\right)
\left(\begin{array}{ccc}2l&2l'&2l''\\0&0&0\end{array}\right)\\
\label{eq:ResolIntegr1D}
&\sum_{j=0}^n\left(\begin{array}{c}l-l'+n-j-1\\n-j\end{array}\right)
\sum_{j'=0}^{n'}\left(\begin{array}{c}l'-l+n'-j'-1\\n'-j'\end{array}\right)
\int_0^1\dd u P_{2l''}(u)Q_{j,j',l+l'}\left(\frac{z^{-1/3}}{4(1-\beta u^2)}\right),\\
\label{eq:ResolIntegr1DQ}
Q_{j,j',l}(x)
=&~\sqrt{\frac{2}{\pi}}
\int_0^{+\infty}\dd r \ee^{-r^2/2}r^{2l+2}
L_j^{(l+1/2)}\left(\frac{r^2}{2}\right)
L_{j'}^{(l+1/2)}\left(\frac{r^2}{2}\right)
\frac{x^3}{r^6/8-x^3}.
\end{align}
The expansion
\begin{align}
\frac{x^3}{r^6/8-x^3} &= \frac{1}{3}\sum_{k=0}^2 \frac{x\ee^{-2i\pi k/3}}{r^2/2-x\ee^{-2i\pi k/3}},
\end{align}
shows that $Q$ can be decomposed as
\begin{align}
\nonumber
Q(x)
=&~ \frac{1}{3} T(x)+ \frac{2}{3}\Real\left(T\left(x\ee^{-2i\pi/3}\right)\right),\\
T_{j,j',l}(x)
=&~\sqrt{\frac{2}{\pi}}
\int_0^{+\infty}\dd r \ee^{-r^2/2}r^{2l+2}
L_j^{(l+1/2)}\left(\frac{r^2}{2}\right)
L_{j'}^{(l+1/2)}\left(\frac{r^2}{2}\right)\frac{x}{r^2/2-x}.
\end{align}
For $\Imag(x)<0$, which is the case for $x=z^{-1/3}/(4(1-\beta u^2))$ and for $x\ee^{-2i\pi/3}$  as $\Imag(z)\rightarrow 0^+$,
\begin{align}
\nonumber
T_{0,0,0}(x)
=&~\sqrt{\frac{2}{\pi}}
\int_0^{+\infty}\dd r 
\frac{x\ee^{-r^2/2}r^2}{r^2/2-x}
=
2x\left[ 1 +  \sqrt{\frac{x}{\pi}}
\int_0^{+\infty}\dd r\ee^{-r^2/2} 
\left(\frac{1}{r-\sqrt{2x}} - \frac{1}{r+\sqrt{2x}} \right)\right]\\
\nonumber
=&
2x\left[1 -i\sqrt{\frac{x}{\pi}}\int_0^{+\infty}\dd t
\int_0^{+\infty}\dd r\ee^{-r^2/2} \ee^{-it\sqrt{2x}}
\left(\ee^{itr}+ \ee^{-itr}\right)\right]\\
\nonumber
=&
2x\left[1 -i\sqrt{2x}\int_0^{+\infty}\dd t\ee^{-t^2/2-it\sqrt{2x}}\right]
=
2x\left[1 -i\sqrt{\pi x}\ee^{-x}\frac{2}{\sqrt{\pi}}\int_{i\sqrt{x}}^{+\infty}\dd t\ee^{-t^2}\right]\\
\label{eq:DerivationResolT000}
=&
2x\left[1 -i\sqrt{\pi x}w(-\sqrt{x})\right]
\end{align}
where $w(z)=\ee^{-z}\erfc(-iz)$ is the Faddeeva function. Using $r^2=2(r^2/2-x+x)$, we then obtain recursively
\begin{align}
T_{0,0,l+1}(x)
=&~\sqrt{\frac{2}{\pi}}
\int_0^{+\infty}\dd r \ee^{-r^2/2}r^{2l+4}
\frac{x}{r^2/2-x}
= 2x \left(T_{0,0,l}(x)+\mathcal{N}_{0,l}^{-2}\right)
= 2x\left(T_{0,0,l}(x)+(2l+1)!!\right).
\end{align}
Using the symmetry $T_{j,j',l}=T_{j',j,l}$, this allows us to construct the family of functions $T_{j,j',l}$ via the recurrence relation of Laguerre polynomials:
\begin{align}
\nonumber
L_{j+1}^{(\alpha)}(y)
=&
\frac{(2j+1+\alpha-y)L_{j}^{(\alpha)}(y)-(j+\alpha)L_{j-1}^{(\alpha)}(y)}{j+1}\\
\nonumber
\Rightarrow
T_{j+1,j',l}(x)
=&
\frac{2j+l+3/2-x}{j+1}T_{j,j',l}(x)
-\frac{j+l+1/2}{j+1}T_{j-1,j',l}(x)\\
\nonumber
&-\frac{x}{j+1}\sqrt{\frac{2}{\pi}}
\int_0^{+\infty}\dd r \ee^{-r^2/2}r^{2l+2}
L_j^{(l+1/2)}\left(\frac{r^2}{2}\right)
L_{j'}^{(l+1/2)}\left(\frac{r^2}{2}\right)\\
\label{eq:RecurrTjjl}
=&~ \frac{2j+l+3/2-x}{j+1}T_{j,j',l}(x)
-\frac{j+l+1/2}{j+1}T_{j-1,j',l}(x)
-\frac{x}{(j+1)\mathcal{N}_{j,l}^2}\delta_{j,j'}.
\end{align}
The explicit calculation of $T_{0,0,0}(x)$ eliminates the apparent divergence of the integrand in Eq.~\ref{eq:ResolIntegr3D} and makes Eq.\ref{eq:ResolIntegr1D} easy to integrate numerically. We thus recover Eq.~7 in the main text.
 
For a spherical cloud with $\beta=0$, the interactions conserve angular momentum $(\Rightarrow l'=l)$, and the integral in Eq.~\ref{eq:ResolIntegr1D} gives $ \delta_{l''}Q_{j,j',2l}(z^{-1/3}/4)$. After simplification,
\begin{align}
\label{eq:ResolvSpherElts}
z\braMket{\Psi_{n_c',n_d',l'}}{\oG(z)}{\Psi_{n_c,n_d,l}}
=&\delta_{n_c,n_c'}\delta_{l,l'}\left(\delta_{n_d,n_d'}
+\mathcal{N}_{n_d,2l}\mathcal{N}_{n_d',2l}
Q_{n_d,n_d',2l}\left(z^{-1/3}/4\right)\right),
\end{align}
giving Eq.~9 in the main text.

\subsection{Characteristic energy scale}
\label{app:EnergyScale}

To obtain the effective Hamiltonian $\oV_e(z_e)=V_0 [z_e -(\oP\oG(z_e)\oP)^{-1}]$ using the matrix elements of $\oG(z)$ calculated above, we need to make an appropriate choice for the dimensionless energy $z_e$. From Eq.~\ref{eq:PropVsResolv}, it is easy to see that if the system's evolution is Markovian, i.e. if $\oP\oU(t)\oP = \exp(-\hat{\Gamma}t - i\hat{\Lambda} t)$, we get $\oV_e=\hat{\Lambda}-i\hat{\Gamma}$ independently on $z_e$. Therefore, the choice of $z_e$ mainly influences our ability to reproduce the non-Markovian part of the evolution. In many problems, $z_e$ is defined as the mean value of $\oV/V_0$, $z_e=\Tr(\oP \oV\oP)/\Tr(V_0\oP)$. In our case, the interactions in the fully symmetric doubly-excited state $\ket{\Psi_{\bf 0}}$ diverge. This effect is non-physical as it arises in a short-range limit where the van-der-Waals approximation is no longer valid. Our approach, providing a finite $\oV_e$, conveniently eliminates it without introducing cutoffs, but this choice of $z_e$ is inappropriate. Instead, as shown in the main text, the brute-force simulations are very well reproduced with $z_e=z_\Omega=\Omega/(2V_0)$ which corresponds to the power-broadened linewidth. In this case, if $\Omega$ depends on time, $\oV_{e}$ will be time-dependent. Alternatively, we can note that the singly-excited qubit state $\ket{R}$ is only laser-coupled to the Gaussian doubly-excited state $\ket{\Psi_{{\bf 0}}}$. Considering that atomic pairs are excited at random positions, which is of course a crude approximation ignoring blockade effects, we obtain a probability distribution for the dimensionless interaction energy $z$ as
\begin{multline}
p(z)= \int\frac{\dd {\bf r}^3}{(2\pi)^{3/2}}\ee^{-r^2/2}\delta\left(\frac{1}{8r^6(1-\beta\cos^2\theta)^3}-z\right) \\
= \sqrt{\frac{2}{\pi}} \int_0^1 \dd u \int_0^{+\infty} \dd r \, r^2 \ee^{-r^2/2}\delta\left(\frac{1}{8r^6(1-\beta u^2)^3}-z\right)
= \frac{1}{12\sqrt{\pi}z^{3/2}}\int_0^1 \frac{\dd u}{(1-\beta u^2)^{3/2}} \ee^{-1/[4z^{1/3}(1-\beta u^2)]}
\end{multline}
which can be integrated using a change of variable $y = \sqrt{1/(1-\beta u^2)-1}$, giving Eq.~10 in the main text:
\begin{align}
p(z)=&~\frac{\ee^{-1/(4z^{1/3})}}{12\sqrt{ \pi \beta}z^{3/2}} \int_0^{\sqrt{\beta/(1-\beta)}} \dd y \ee^{-y^2/(4z^{1/3})}
= \frac{\ee^{-1/(4z^{1/3})}}{12\sqrt{\beta}z^{4/3}} \erf\left(\sqrt{\frac{\beta}{1-\beta}}\frac{1}{2z^{1/6}}\right).
\end{align}
In the limit $\beta\rightarrow 0$ of a spherical cloud, this reduces to $p(z)=\ee^{-1/(4z^{1/3})}/(12\sqrt{\pi}z^{3/2})$ which has a maximum in $z_0=18^{-3}\approx 1.7\times 10^{-4}$. This value changes little when the cloud flattens along the $z$ axis: for $\beta \rightarrow 1$ ($\sigma_z/\sigma\rightarrow 0$), this maximum is in $16^{-3}\approx 2.4\times 10^{-4}$. In the opposite, prolate regime, the limit $\beta\rightarrow -\infty$ of an infinitely long cloud is irrelevant as the blockade vanishes in this case.  As shown in Fig.~2 in the main text, choosing $z_e=z_0$ provides a good approximation of the brute-force simulations when the blockade is not too weak, and makes $\oV_e$ time-independent.

One should note that $\oV_e$ would be totally independent on $z_e$ if the system's evolution in the subspace $\{\oP\}$ was Markovian. In this case, the forward propagator in Eq.~\ref{eq:PropRestrict} could be written as $\oP\oU(t)\oP=\exp(-i\oH_P t)$ with $\oH_P$ a non-Hermitian Hamiltonian. The resolvent would become $\oP\oG(z)\oP=1/(z-\oV_p)$, giving $\oV_e=\oH_P$ independently on $z$. Therefore, the choice of $z_e$ primarily influences the model's ability to reproduce the non-Markovian part of the evolution. When the blockade is not too weak, it can be considered as a higher-order effect, making the model weakly sensitive to the choice of $z_e$ and allowing one to set $z_e=z_0$ for convenience. 

\subsection{Spherical cloud}

\label{app:EffInteractSpher}

For a spherical cloud the effective interaction potential $\oV_e=\hat{\Lambda}-i\hat{\Gamma}$ is block-diagonal in $n_c$ and $l$ (see Eq.~\ref{eq:ResolvSpherElts}). We show here that inside each of these blocks the decay matrix $\hat{\Gamma}$ has a single non-zero eigenvalue, allowing one to use a single collapse operator per $(n_c,l)$ block in the master equation. From Eqs.~\ref{eq:ResolvSpherElts}, \ref{eq:ResolIntegr1D} and \ref{eq:ResolIntegr3D}, 
\begin{equation}
\braMket{\Psi_{n_c,n_d',l}}{\oG(z)}{\Psi_{n_c,n_d,l}}
= \int_0^{+\infty} \dd r
r^2
\frac{R_{n_d,2l}(r)R_{n_d',2l}(r)}{z-1/(8r^6)}.
\end{equation}
In the limit $\Imag(z)\rightarrow 0^+$,
\begin{align}
\nonumber
\braMket{\Psi_{n_c,n_d',l}}{\oG(z)}{\Psi_{n_c,n_d,l}}
=&
\int_0^{+\infty} \dd r
r^2 R_{n_d,2l}(r)R_{n_d',2l}(r)
\left(\mathcal{P}\frac{1}{z-1/(8r^6)}-i\pi\delta\left(z-\frac{1}{8r^6}\right)\right)\\
\nonumber
=&~ \mathcal{P} \int_0^{+\infty} \dd r
r^2 
\frac{R_{n_d,2l}(r)R_{n_d',2l}(r)}{z-1/(8r^6)}
-i\frac{4\pi r_c^9}{3}R_{n_d,2l}(r_c)R_{n_d',2l}(r_c),\\
r_c =&~ \frac{1}{\sqrt{2} z^{1/6}},
\end{align}
where $\mathcal{P}$ stands for the Cauchy principal value. For given $n_c$ and $l$, this allows us to rewrite $ \oG(z)=\oG_r(z)+i\oG_i(z)$ where $\oG_r=\Real(\oG(z))$ and $\oG_i=\Imag(\oG(z))$ are both Hermitian, with
\begin{align}
\oG_r(z)
=&~ \sum_{n_d,n_d'} \mathcal{P} \int_0^{+\infty} \dd r r^2 
\frac{R_{n_d,2l}(r)R_{n_d',2l}(r)}{z-1/(8r^6)} \ketbra{\Psi_{n_c,n_d',l}}{\Psi_{n_c,n_d,l}},\\
\nonumber
\oG_i(z)
=&~ 
 \alpha_{n_c,l} \ketbra{\chi_{n_c,l}}{\chi_{n_c,l}},\\
\alpha_{n_c,l}
\nonumber
=&~ - \frac{4\pi r_c^9}{3}\sum_{n_d} R_{n_d,l}^2(r_c),\\
\ket{\chi_{n_c,l}}
=&~
\frac{1}{\sqrt{ \sum_{n_d} R_{n_d,l}^2(r_c)}}
\sum_{n_d}R_{n_d,l}(r_c)\ket{\Psi_{n_c,n_d,l}}.
\end{align}
where the sums over $n_d$ run from $0$ to $n_{d,max}=n_{max}-n_c-2l$. By design, $\oG$ has an inverse  $\oG^{-1}=z -\hat{\Lambda}/V_0 +i\hat{\Gamma}/V_0$ such that, still in this ($n_c,l$) subspace,
\begin{equation}
\label{eq:InvGtimesG}
V_0 \oG^{-1}\oG = V_0 \hat{1}= (V_0 z -\hat{\Lambda})\oG_r-\hat{\Gamma}\oG_i+i\left((V_0 z -\hat{\Lambda})\oG_i+\hat{\Gamma}\oG_r\right),
\end{equation}
Defining the projector $\oP_\perp= \hat{1}-\ketbra{\chi_{n_c,l}}{\chi_{n_c,l}}$, the imaginary part of this equation gives $\hat{\Gamma}\oG_r\oP_\perp=0$. As $\det(\oG)=\det(\oG_r)+i\alpha_{n_c,l}\det(\oP_\perp\oG_r\oP_\perp)\neq 0$, $\rank(\oG_r\oP_\perp)=\Tr(\oP_\perp)=\Tr( \hat{1})-1$. Since  $\hat{\Gamma}$ annihilates a subspace of dimension  $\Tr( \hat{1})-1$, it can be written as $\hat{\Gamma}=\Gamma_{n_c,l}\ketbra{\zeta_{n_c,l}}{\zeta_{n_c,l}}$ where $\oP_\perp\oG_r\ket{\zeta_{n_c,l}}=0$, which concludes the proof.

When $\oG_r(z)$ is invertible, which is the case except for a finite number of $z$ values,
\begin{equation}
\ket{\zeta_{n_c,l}} = \frac{\oG_r^{-1} \ket{\chi_{n_c,l}}}{\sqrt{ \bra{\chi_{n_c,l}}\oG_r^{-2}\ket{\chi_{n_c,l}}}}.
\end{equation}
To find the decay rate $\Gamma_{n_c,l}$, as well as $\hat{\Lambda}$, we can use the real and imaginary parts of Eq.~\ref{eq:InvGtimesG} to express $\bra{\chi_{n_c,l}}{V_0 z-\hat{\Lambda}}\ket{\chi_{n_c,l}}$ in two different ways:
\begin{align}
\nonumber
\bra{\chi_{n_c,l}}{V_0 z-\hat{\Lambda}}\ket{\chi_{n_c,l}}
=&~ \braMket{\chi_{n_c,l}}{V_0\oG_r^{-1} +  \hat{\Gamma}\oG_i\oG_r^{-1}}{\chi_{n_c,l}}
=
V_0 \bra{\chi_{n_c,l}}{\oG_r^{-1}}\ket{\chi_{n_c,l}} + \alpha_{n_c,l} \Gamma_{n_c,l} \frac{\bra{\chi_{n_c,l}}{\oG_r^{-1}}\ket{\chi_{n_c,l}}^3}{\bra{\chi_{n_c,l}}{\oG_r^{-2}}\ket{\chi_{n_c,l}}},\\
\nonumber
\bra{\chi_{n_c,l}}{V_0 z-\hat{\Lambda}}\ket{\chi_{n_c,l}}
=&~ -\frac{1}{\alpha_{n_c,l}}\bra{\chi_{n_c,l}}{\hat{\Gamma}\oG_r}\ket{\chi_{n_c,l}}
= -\frac{\Gamma_{n_c,l}}{\alpha_{n_c,l}} \frac{\bra{\chi_{n_c,l}}{\oG_r^{-1}}\ket{\chi_{n_c,l}}}{\bra{\chi_{n_c,l}}{\oG_r^{-2}}\ket{\chi_{n_c,l}}}\\
\Rightarrow \Gamma_{n_c,l}
=&~
- V_0 \frac{\alpha_l \bra{\chi_{n_c,l}}{\oG_r^{-2}}\ket{\chi_{n_c,l}}}{1+\alpha_l^2 \langle \chi_{n_c,l} | \oG_r^{-1} | \chi_{n_c,l}\rangle^2}\\
\Rightarrow 
\hat{\Lambda}=&~V_0(z-\oG_r^{-1})-\alpha_{n_c,l} \langle \chi_{n_c,l} | \oG_r^{-1} | \chi_{n_c,l}\rangle \hat{\Gamma}.
\end{align}

\subsection{Resonant dipole-dipole interactions}
\label{app:ResDipDipInt}

This formalism can be transposed from van-der-Waals to other types of interactions. In particular, by using a static electric filed, one can tune Rydberg atoms to a F\"orster resonance where the doubly-excited Rydberg state $|r,r\rangle$ of atoms $a$ and $b$ is resonantly coupled to an exchange-symmetric state $|r',r''\rangle$ involving two other Rydberg states $r'$ and $r''$ with dipole-allowed transitions to $r$ and opposite energy shifts. The interaction potential is in this case
\begin{equation}
V = \frac{1}{4\pi\epsilon_0}\frac{{\bf d}_a\cdot{\bf d}_b-3({\bf d}_a\cdot{\bf n})({\bf d}_b\cdot{\bf n})}{R^3}
\end{equation}
where ${\bf d}_i$ is the electric dipole of atom $i$ on the considered transition, $R=|{\bf R}_a-{\bf R}_b|$ the distance between the two atoms, and ${\bf n}=({\bf R}_a-{\bf R}_b)/R$ the unit vector of the interatomic axis. For simplicity, we consider that the transitions $r-r'$ and $r-r''$ are $\pi$-polarized, i.e. that the two dipoles are aligned with the $z$ axis of the cloud. Other polarizations lead to more cumbersome expressions but can be treated similarly. Defining, like previously, $C_3 = d^2/(4\pi\epsilon_0)$, $V_0 = C_3/\sigma^3$, ${\bf r}=(X/\sigma,Y/\sigma,Z/\sigma_z)$, ${\bf r}_c=({\bf r}_a+{\bf r}_b)/\sqrt{2}$ and ${\bf r}_d=({\bf r}_a-{\bf r}_b)/\sqrt{2}$,
\begin{multline}
V = C_3 \frac{1-3\cos^2(\theta)}{R^3} = C_3 \frac{(X_b-X_a)^2+(Y_b-Y_a)^2-2(Z_b-Z_a)^2}{((X_b-X_a)^2+(Y_b-Y_a)^2+(Z_b-Z_a)^2)^{5/2}}
=\frac{V_0}{2\sqrt{2}}\frac{x_d^2+y_d^2-2(1-\beta)z_d^2}{(x_d^2+y_d^2+(1-\beta)z_d^2)^{5/2}}\\
=\frac{V_0}{2\sqrt{2} r_d^3}\frac{1-(3-2\beta)\cos^2(\theta_d)}{(1-\beta\cos^2(\theta_d))^{5/2}}.
\end{multline}
We work in a frozen gas approximation where mechanical effects are neglected. Therefore, the sign of $C_3$, defining whether atoms attract or repel each other for a given direction $\theta$, plays no role. In the following we take $C_3>0$ for simplicity. With our normalized coordinates, this potential has the same symmetries as the van-der-Waals potential for an aspherical cloud. Its eigenstates are of the form $(|r,r\rangle\pm|r',r''\rangle)/\sqrt{2}$, with corresponding eigenenergies $\pm V$. As a result, we now have two doubly-excited subspaces labeled by $+$ and $-$, spanned by
\begin{align}
\nonumber
|\Psi_{n_c,l_c,n_d,l_d,m}^\pm\rangle
=&~ \frac{|\Phi_{n_c,2l_c,-m,n_d,2l_d,m}^\pm\rangle+|\Phi_{n_c,2l_c,m,n_d,2l_d,-m}^\pm\rangle}{\sqrt{2(1+\delta_m)}},\\
\nonumber
|\Phi_{n_c,l_c,m_c,n_d,l_d,m_d}^\pm\rangle
=&~ \int \dd{\bf r}_a^3 \dd{\bf r}_b^3 \psi_{n_c,l_c,m_c}({\bf r}_c)\psi_{n_d,l_d,m_d}({\bf r}_d)\\
&~ \frac{1}{2}\left(\os_{rg}({\bf r}_a)\os_{rg}({\bf r}_b)\pm \frac{\os_{r'g}({\bf r}_a)\os_{r''g}({\bf r}_b)+\os_{r''g}({\bf r}_a)\os_{r'g}({\bf r}_b)}{\sqrt{2}}\right)\ket{G}.
\end{align}
These subspaces are uncoupled from each other by the interactions, and laser-coupled to the singly-excited subspace via the matrix elements already defined in Supp. Mat.~\ref{app:DblExcDecomp} and in 
equation 4 in the main text, up to a factor $1/\sqrt{2}$:
\begin{equation}
\oSd
= \ket{\psi_{0,0}}\bra{G} + \sum_{n_d,l_c,l_d,m\geq 0} S_{n,l;l_c,n_d,l_d,m} \frac{|\Psi_{n_c,l_c,n_d,l_d,m}^+\rangle +|\Psi_{n_c,l_c,n_d,l_d,m}^-\rangle }{\sqrt{2}}\bra{\psi_{n,l}}.
\end{equation}
By symmetry, they will have opposite interaction-induced frequency shifts and identical decays: $\oV_e^\pm = \pm\hat{\Lambda}-i\hat{\Gamma}$. In the following, we will focus on $\oV_e^+$.

\begin{figure}
\centering
	\includegraphics[width=170mm]{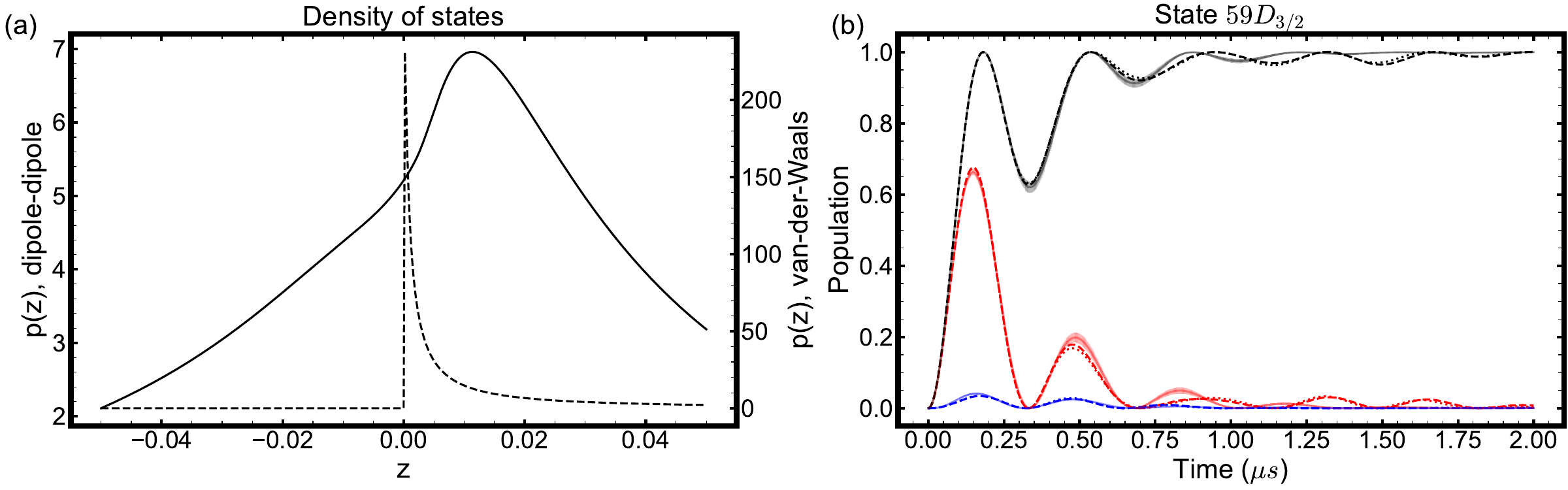}
	\caption{(a) Density of states for dipole-dipole interactions given by Eq.~\ref{eq:DensStatesDip} (solid), compared to van-der-Waals (dashed). Vertical scales are different. (b) Populations of the qubit's excited state $\ket{R}$ (red), of the doubly-excited symmetric state $\ket{\Psi_{0,0,0}}$ (blue), and of all excited states (black), for the Förster-resonant state $59D_{3/2}$ where interactions are of the dipole-dipole type. A brute-force simulation (full lines) is compared with our model where $z_e=z_\Omega$ (dashed) or $z_e=z_0$ (dotted). The shaded areas around the full lines correspond to the standard deviations obtained from $8$ brute-force simulations. To speed up the latter, doubly-excited states with interaction energies exceeding $32\times\Omega/2$ are considered as far off-resonant and discarded from the Hilbert space.}
	\label{fig:ModelVsBruteForceDip}
\end{figure}

Following Appendix \ref{app:EnergyScale}, the dimensionless density of doubly-excited states in the $+$ subspace is given by:
\begin{align}
\nonumber
p(z) =&~ \int\frac{\dd {\bf r}^3}{(2\pi)^{3/2}}\ee^{-r^2/2}\delta\left(\frac{1}{2\sqrt{2} r^3}\frac{1-(3-2\beta)\cos^2(\theta)}{(1-\beta\cos^2(\theta))^{5/2}}-z\right) = \sqrt{\frac{2}{\pi}} \int_0^1 \dd u \int_0^{+\infty} \dd r \, r^2 \ee^{-r^2/2}\delta\left(\frac{f(u)}{ r^3}-z\right),\\
\nonumber
f(u) =&~ \frac{1-(3-2\beta)u^2}{2\sqrt{2}(1-\beta u^2)^{5/2}};\\
\nonumber
p(z>0) =&~ \sqrt{\frac{2}{\pi}} \int_0^{1/\sqrt{3-2\beta}} \dd u \frac{f(u)}{3 z^2} \ee^{-(f(u)/z)^{2/3}/2},\\
\nonumber
p(z<0) =&~ -\sqrt{\frac{2}{\pi}} \int_{1/\sqrt{3-2\beta}}^1 \dd u \frac{f(u)}{3 z^2} \ee^{-(f(u)/z)^{2/3}/2},\\
\label{eq:DensStatesDip}
p(0) =&~ \sqrt{\frac{2}{\pi}} \int_0^{+\infty} \dd r \, \frac{r^5}{|f'(1/\sqrt{3-2\beta})|} \ee^{-r^2/2}
=\frac{16}{\sqrt{3\pi}}\frac{(1-\beta)^{5/2}}{(1-2\beta/3)^3}.
\end{align}
The remaining integrals are calculated numerically. As visible in Fig.~\ref{fig:ModelVsBruteForceDip}(a), this density of states is much less peaked than for the van-der-Waals case. For $\beta=0$, it reaches its maximum in $z_0\approx 1.13 \times 10^{-2}$, which gives an effective cloud's size $\sigma/z_0^{1/3}\approx 4.46 \sigma$ very close to that obtained for van-der-Waals interactions ($3\sqrt{2}\sigma\approx 4.24\sigma$). In what follows, we will assume $\Real(z)\geq 0$.

To calculate the resolvent in the $+$ subspace, let us write
\begin{align}
\nonumber
zG(z,{\bf r}) =&~ \frac{z}{z-f(\cos^2(\theta))/r^3} = 1+\frac{ f(\cos^2(\theta))/(2^{3/2}z)}{(r/\sqrt{2})^3-f(\cos^2(\theta))/(2^{3/2}z)} = 1+\frac{1}{3}\sum_{k=0}^2\frac{g(\cos^2(\theta))\ee^{2i\pi k/3}}{r/\sqrt{2}-g(\cos^2(\theta))\ee^{2i\pi k/3}},\\
\label{eq:ResolDDYofU}
g(u) =&~ \frac{|f(u)|^{1/3}\sign(f(u))}{\sqrt{2}z^{1/3}}
= \frac{|1-(3-2\beta)u^2|^{1/3}\sign(1-(3-2\beta)u^2)}{2z^{1/3}(1-\beta u^2)^{5/6}}.
\end{align}
This choice for $g$ gives $\Imag(g^2(u))<0$ which simplifies following calculations. The equation giving the matrix elements $G_{n,l;n',l';m}(z)$ is the same as Eq.~\ref{eq:ResolIntegr1D} with a different function $Q$:
\begin{align}
\nonumber
z G_{n,l;n',l';m}(z)
=&~
 \delta_{n,n'}\delta_{l,l'}
+(-1)^m\mathcal{N}_{n,2l}\mathcal{N}_{n',2l'}\sqrt{(4l+1)(4l'+1)}\sum_{l''=|l-l'|}^{l+l'}(4l''+1)
\left(\begin{array}{ccc}2l&2l'&2l''\\-m&m&0\end{array}\right)
\left(\begin{array}{ccc}2l&2l'&2l''\\0&0&0\end{array}\right)\\
\nonumber
&\sum_{j=0}^n\left(\begin{array}{c}l-l'+n-j-1\\n-j\end{array}\right)
\sum_{j'=0}^{n'}\left(\begin{array}{c}l'-l+n'-j'-1\\n'-j'\end{array}\right)
\int_0^1\dd u P_{2l''}(u) Q_{j,j',l+l'}\left(g(u)\right),\\
\nonumber
Q_{j,j',l}(x)
=&~
\sqrt{\frac{2}{\pi}}
\int_0^{+\infty}\dd r \ee^{-r^2/2}r^{2l+2}
L_j^{(l+1/2)}\left(\frac{r^2}{2}\right)
L_{j'}^{(l+1/2)}\left(\frac{r^2}{2}\right)
\frac{x^3}{(r/\sqrt{2})^3-x^3} \\
\nonumber
=&~
 \frac{1}{3} T_{j,j',l}(x)+ \frac{2}{3}\Real\left(T_{j,j',l}\left(x\ee^{2i\pi/3}\right)\right),\\
T_{j,j',l}(x)
=&~ \sqrt{\frac{2}{\pi}}
\int_0^{+\infty}\dd r \ee^{-r^2/2}r^{2l+2}
L_j^{(l+1/2)}\left(\frac{r^2}{2}\right)
L_{j'}^{(l+1/2)}\left(\frac{r^2}{2}\right)\frac{x}{r/\sqrt{2}-x}.
\end{align}
As we assumed  $\Imag(z)\rightarrow 0^+$ and $\Real(z)>0$, $\Imag(x^2)<0$ for $x=g(u)$ and $x=g(u)\ee^{2i\pi/3}$. This allows us to reuse the derivation of Eq.~\ref{eq:DerivationResolT000} to obtain
\begin{align}
\nonumber
T_{0,0,0}(x)
=&~\sqrt{\frac{2}{\pi}}
\int_0^{+\infty}\dd r 
\frac{x\ee^{-r^2/2}r^2}{r/\sqrt{2}-x}
= \sqrt{\frac{2}{\pi}} \int_0^{+\infty}\dd r 
\frac{x^2\ee^{-r^2/2}r^2}{r^2/2-x^2}
+ \frac{x}{\sqrt{\pi}}\int_0^{+\infty}\dd r 
\frac{r^3 \ee^{-r^2/2}}{r^2/2-x^2}
\\
\nonumber
=&~
2x^2\left[1 -i\sqrt{\pi x^2}w\left(-\sqrt{x^2}\right)\right]
+ \frac{2x}{\sqrt{\pi}}\int_0^{+\infty}\dd r \,r\,\ee^{-r^2/2}
\left(1+\frac{x^2}{r^2/2-x^2}\right)
\\
\nonumber
=&~
2x^2\left[1 -i\sqrt{\pi x^2}w\left(-\sqrt{x^2}\right)\right]
+ \frac{2x}{\sqrt{\pi}}
\left(1+x^2\ee^{-x^2}\int_{-x^2}^{+\infty}\frac{\ee^{-t}}{t}\dd t\right)
\\
=&~
2x^2\left[1 -i\sqrt{\pi x^2}w\left(-\sqrt{x^2}\right)\right]
+ \frac{2x}{\sqrt{\pi}}
\left(1-x^2\ee^{-x^2}\left(i\pi+\Ei(x^2)\right)\right).
\end{align}
Using $(r^2/2)/(r/\sqrt{2}-x)=x^2/(r/\sqrt{2}-x)+x+r/\sqrt{2}$, we proceed by recurrence to get
\begin{align}
\nonumber
T_{0,0,l+1}(x)
=&~\sqrt{\frac{2}{\pi}}
\int_0^{+\infty}\dd r \ee^{-r^2/2}
\frac{r^{2l+4} x}{r/\sqrt{2}-x}\\
\nonumber
=&~ 2x^2 T_{0,0,l}(x)
+ 2x^2\sqrt{\frac{2}{\pi}} \int_0^{+\infty}\dd r\, r^{2l+2} \ee^{-r^2/2}
+ \frac{2x}{\sqrt{\pi}} \int_0^{+\infty}\dd r \,r^{2l+3} \ee^{-r^2/2}\\
=&~ 2x^2(T_{0,0,l}(x) + (2l+1)!!) + 2x\frac{2^{l+1}(l+1)!}{\sqrt{\pi}}.
\end{align}
The recurrence for the generic term given by Eq.~\ref{eq:RecurrTjjl} becomes
\begin{align}
\nonumber
T_{j+1,j',l}(x)
=&
\frac{2j+l+3/2-x^2}{j+1}T_{j,j',l}(x)
-\frac{j+l+1/2}{j+1}T_{j-1,j',l}(x)\\
\nonumber
&-\frac{x^2}{j+1}\sqrt{\frac{2}{\pi}}
\int_0^{+\infty}\dd r \ee^{-r^2/2}r^{2l+2}
L_j^{(l+1/2)}\left(\frac{r^2}{2}\right)
L_{j'}^{(l+1/2)}\left(\frac{r^2}{2}\right)\\
\nonumber
&-\frac{x}{j+1}\frac{1}{\sqrt{\pi}}
\int_0^{+\infty}\dd r \ee^{-r^2/2}r^{2l+3}
L_j^{(l+1/2)}\left(\frac{r^2}{2}\right)
L_{j'}^{(l+1/2)}\left(\frac{r^2}{2}\right)\\
=&~ \frac{2j+l+3/2-x^2}{j+1}T_{j,j',l}(x)
-\frac{j+l+1/2}{j+1}T_{j-1,j',l}(x)
-\frac{x^2}{(j+1)\mathcal{N}_{j,l}^2}\delta_{j,j'}
-\frac{x}{j+1}\frac{2^{l+1}}{\sqrt{\pi}}v_{j,j',l},\\
\nonumber
v_{j,j',l} 
=&~ \int_0^{+\infty}\dd t \ee^{-t}t^{l+1}
L_j^{(l+1/2)}(t)L_{j'}^{(l+1/2)}(t)\\
\nonumber
=&~ \sum_{i=0}^{j}\sum_{i'=0}^{j'}
\left(\begin{array}{c} j-i-3/2 \\ j-i\end{array}\right)
\left(\begin{array}{c} j'-i'-3/2 \\ j'-i'\end{array}\right)\int_0^{+\infty}\dd t \ee^{-t}t^{l+1}
L_i^{(l+1)}(t)L_{i'}^{(l+1)}(t)\\
\nonumber
=&~ \sum_{i=0}^{\min(j,j')}
\left(\begin{array}{c} j-i-3/2 \\ j-i\end{array}\right)
\left(\begin{array}{c} j'-i-3/2 \\ j'-i\end{array}\right)
\frac{(i+l+1)!}{i!}
\end{align}
where, in the calculation of $v_{j,j',l}$, we used Eq.~\ref{EqLagNAlphaLagIBeta}.

Figure~\ref{fig:ModelVsBruteForceDip}(b) compares the result of this model with brute force simulations for the state $|r,r\rangle=|59D_{3/2},59D_{3/2}\rangle$ of $^{87}$Rb, Förster-resonant with $|r',r''\rangle=|57F_{5/2},61P_{3/2}\rangle$, in a spherically-symmetric cloud with a Gaussian radius $\sigma=\unit{3.5}{\micro\meter}$. Like for the van-der-Waals case, we take $\Omega=2\pi\times\unit{3}{\mega\hertz}$, which gives $z_\Omega=\Omega/(2V_0)=9.2\times 10^{-3}=0.82 z_0$. The brute force simulations consider $200$ atoms; the model includes states with $n_{max}=4$.

The agreement between the two approaches is quite satisfactory even though the blockade is bad, resembling that of the 95S state shown in Fig.~2 of the main paper. This resemblance is not accidental. When driving the cloud with a Rabi frequency $\Omega$, one typically populates doubly-excited states with $z\in[-z_\Omega,z_\Omega]$. The blockade imperfection is thus characterized by $\epsilon\sim \int_{-z_\Omega}^{z_\Omega}p(z)\dd z$. For the simulations presented here, $\epsilon=10\%$, close to the $14\%$ of the 95S state in Fig.~2.

\section{Experimental Protocol}
\label{app:ExptProtocol}

This Supplemental Material presents additional details of our experimental protocol. More information can be found in references \cite{Vaneecloo2022} and \cite{Covolo2025}.

We prepare a small cloud of cold $^{87}$Rb atoms in \unit{100}{\milli\second} through a sequence of cooling, transport and loading into a crossed dipole trap (CDT), as described in detail in Ref.~\cite{Vaneecloo2022}. Within the CDT, we find $N \approx 800$ atoms at a temperature of \unit{3}{\micro\kelvin}, distributed with a Gaussian profile of rms radius $\sigma = ((2\sigma_\rho^2 + \sigma_z^2)/3)^{1/2} = \unit{5.20(2)}{\micro\meter}$, where $\sigma_\rho = \unit{5.62(2)}{\micro\meter}$ is the radius in the plane of the CDT beams and $\sigma_z = \unit{4.22(2)}{\micro\meter}$ is the radius along the orthogonal axis. The cloud is coupled to the Gaussian optical mode of a single-ended running-wave medium-finesse cavity with a waist $w = \unit{21}{\micro\meter}$ and a field decay rate $\kappa = 2 \pi \times \unit{2.89}{\mega\hertz}$.
The relevant atomic states in the system are the ground state $g : \lbrace 5S_{1/2}, F = 1, m_F = 1 \rbrace$, the Rydberg state $r : \lbrace nS_{1/2}, J = 1/2, m_J = 1/2, I = 3/2, m_I = 3/2  \rbrace$ where $n\in\{80,95,109\}$, and the intermediate state $e : \lbrace 5P_{1/2}, F = 2, m_F = 2 \rbrace$ which has a dipole decay rate of $\gamma = 2\pi \times \unit{2.87}{\mega\hertz}$.

The main part of the experimental protocol, where Rydberg blockade imperfections play a major role, consists in driving a two-photon transition from $g$ to $r$ with a pair of laser beams. As described in the main text, for perfect Rydberg blockade the system should oscillate between the ground state $\ket{G}$ and the singly-excited symmetric Rydberg state $\ket{R}$ with a collective Rabi frequency $\Omega = \sqrt{N_0} \Omega_1 \Omega_2 / 2\Delta$, where $N_0 = N/(1+4\sigma_\rho^2/w^2) \sim 620$ is the effective number of atoms coupled to the cavity mode, $\Omega_1$ and $\Omega_2$ are the respective Rabi frequencies on the $g-e$ and $e-r$ transitions, and  $\Delta = - 2\pi \times \unit{500}{\mega\hertz}$ is the detuning from the state $e$. Blockade imperfections create a leakage into a large manifold of asymmetric Rydberg states. This drive sequence lasts $t_d=\unit{500}{\nano\second}$, shaped by the temporal profile of $\Omega_1$ encompassed by $\Omega_2$ as shown on the left graph of Fig.~3(c) in the main text. The rotation angle $\Omega t_d$ of the qubit is controlled by setting $\Omega_2 = 2\pi \times \unit{7.8(1)}{\mega\hertz}$ and adjusting $\Omega_1$ between 0 and $2\pi \times \unit{22}{\mega\hertz}$.

Once the driving sequence is completed, we perform one of the two following measurements:
\begin{itemize}
	\item G/not(G):  To detect if the qubit is in $\ket{G}$, we follow the protocol described in detail in Ref.~\cite{Vaneecloo2022} and measure the transmission of a weak probe beam, resonant with the cavity and with the $g-e$ transition, in presence of a control beam resonantly coupling $e$ with the Rydberg state $r'=78S_{1/2}$ (Fig.~3(b) in the main text). 
	
	When the qubit is in \ket{G}, the control beam opens an electromagnetically induced transparency (EIT) window for the probe photons, transmitted through the cloud as dark Rydberg polaritons. Keeping the probe power sufficiently low to avoid Rydberg blockade between these polaritons, we then detect a constant photon flux $\phi_G$ at the output of the cavity. The probability to detect $n$ photon during an integration time $t_i$ is then given by a Poissonian law $P_G(n) = \mathcal{P}(n, t_i \phi_G ) = e^{-t_i \phi_G}(t_i \phi_G)^n / n!$.
	
	On the other hand, when the cloud contains a Rydberg excitation, the blockade between $r$ and $r'$ suppresses the EIT and the transmitted photon flux drops to a residual value $\phi_r$. Eventually, the Rydberg excitation decays and the photon flux $\phi_G$ is restored. As shown in Ref.~\cite{Vaneecloo2022} appendix C, the probability to detect $n$ photons during a time $t_i$, accounting for the finite Rydberg lifetime $\tau_r$, is then
	\begin{equation}
		P_R(n) = \mathcal{P}(n, t_i \phi_r) e^{-t_i/\tau_r} + \int_0^{t_i} \mathcal{P}(n, t\phi_r + (t_i - t) \phi_G) e^{-t/\tau_r} \frac{\mathrm{d}t}{\tau_r}
	\end{equation}
	
	 In practice, the detection step begins \unit{5.01}{\micro\second} after the driving sequence. For given experimental settings, we reconstruct the distribution $P(n)$ of photon counts integrated during  $t_i = \unit{45}{\micro\second}$, fit it with the function $\eta_r P_r(n) + (1-\eta_r) P_G(n)$, and retrieve the Rydberg population $\eta_r=1-\Tr(\orho\ketbra{G}{G})$ from this fit, together with $\phi_r$. The best-fitted values $\eta_r$ are plotted in Fig.~3(b) in the main text, with a correction factor 1.012 accounting for the decay during the \unit{5.01}{\micro\second} between the drive pulse and the beginning of the detection.
	 
	\item R/not(R):  To estimate the population in the qubit state $\ket{R}$, we increase the Rabi frequency $\Omega_m$ of the mapping beam, resonant with the transition $e-r$, from $0$ to $2 \pi \times \unit{7.4(1)}{\mega\hertz}$ (see cyan arrows and curve in Fig.~3(c)). As detailed in \cite{Magro2023}, the strong collective coupling $g = 2\pi \times \unit{9.70(6)}{\mega\hertz}$ between the atoms and the cavity mode ($g \gg \gamma,\kappa$) enables an efficient mapping of $\ket{R}$ to an intracavity photon, which is then detected by a single-photon detector (SPD) at the cavity output.

\end{itemize}

\section{Numerical simulations}

\label{app:Simul}
This Supplemental Material provides details about the numerical simulations of the experimental results in Fig.~3. We perform $2\times 3\times 37$ simulation runs corresponding to two types of measurements (G/not(G) and R/not(R)), three Rydberg states (80S, 95S and 109S), and 37 different values of the Rabi frequency $\Omega_1$ on the lower branch of the two-photon transition, ranging from 0 to $2\pi \times \unit{22}{\mega\hertz}$. We neglect the weak ellipticity of our cloud and, to reduce the Hilbert space, consider it as spherically symmetric with a Gaussian r.m.s radius $\sigma$ defined in Supp. Mat.~\ref{app:ExptProtocol}. We verified that this assumption has little effect on the precision of the simulations given our experimental uncertainties, but it makes them much faster. We adapt the basis size to the chosen Rydberg state, making it larger when the blockade is weaker and asymmetric states are populated faster: we set $n_{max}=(3,6,8)$ for the states $(109,95,80)S_{1/2}$, respectively. The basis then contains $n_{max}+1$ singly-excited states \ket{\psi_n} and 
\begin{equation}
\sum_{n=0}^{n_{max}}\sum_{l=0}^{\lfloor n_{max}/2\rfloor}(n-2l+1)
=\left(\left\lfloor\frac{n_{max}}{2}\right\rfloor+1\right)
\left(\left\lfloor\frac{n_{max}}{2}\right\rfloor+2\right)
\left(n_{max}-\frac{4}{3}\left\lfloor\frac{n_{max}}{2}\right\rfloor+\frac{1}{2}\right)
\end{equation}  
doubly-excited states $\ket{\Psi_{n_c,n_d,l}}$ with $n_c+n_d+2l=n\leq n_{max}$. The integration of the master equation is performed on a desktop computer using the Python package QuTiP \cite{qutip2}. The solver supports time-dependent Rabi frequencies, with temporal profiles obtained by fitting the experimental pulses to theoretical curves (see Fig.~3(c) in the main text). To keep the Lamb-shifted interaction matrix $\Lambda$ and the decay matrix $\Gamma$ time-independent, we set the effective energy to $z_e=z_0$. We simulate the total Rydberg population (G/non(G) measurement) and the number of extracted photons (R/not(R) measurement) with two slightly different versions of the master equation, described below.

\subsection{G/not(G) measurement: simulating the total Rydberg population}

For the G/not(G) measurement, which estimates the probability for at least one Rydberg excitation to be present in the cloud, we simulate only the driving part of the experimental protocol. Errors occurring during the EIT probing, related to the finite Rydberg lifetime and to the residual transmission in the not(G) case, are accounted for via a semiclassical model described in Supp. Mat.~\ref{app:ExptProtocol} and in Ref.~\cite{Vaneecloo2022} Appendix C.

As the two-photon driving, with an effective time-dependent Rabi frequency $\Omega = \sqrt{N_0} \Omega_1 \Omega_2 / 2\Delta$, is far off-resonant from the intermediate atomic state $\ket{e}$, we restrict our Hilbert space to the collective states introduced previously (the ground state $\ket{G}$, the singly- and doubly-excited manifolds  $\{\ket{\psi_n}\}$ and $\{\ket{\Psi_{n_c,n_d,l}}\}$ with $n_c+n_d+2l=n\leq n_{max}$ and the doubly-excited ``continuum'' state $\ket{C}$ populated by the interactions), to which we add a singly-excited ``continuum'' state \ket{c} to account for thermal dephasing as explained below. 

In addition to the laser driving and to the Rydberg interactions, described by the operators $\oS$, $\hat{\Lambda}$ and $\hat{\Gamma}$, the simulations include two experimental imperfections. We neglect the $\sim \unit{100}{\micro\second}$ lifetime of Rydberg excitations, much longer than the experimental timescale $\sim  \unit{2}{\micro\second}$, but we include the pure dephasing rate $\gamma_r = 2\pi \times \unit{40}{\kilo\hertz}$ between the Rydberg manifold and the ground state, measured in Ref.~\cite{Magro2023}. In addition, the finite temperature $T$ of the cloud leads to a Maxwell-Boltzmann distribution of atomic velocities. This translates into a Gaussian distribution $\ee^{-\omega^2/(2\omega_0^2)}/\sqrt{2\pi\omega_0^2}$ of Doppler shifts $\omega$ with a width $\omega_0 = k\sqrt{k_B T/m}= 2\pi \times \unit{57}{\kilo\hertz}$ where $k$ is the spin wave vector, $k_B$ is the Boltzmann constant and $m$ is the atomic mass. This inhomogeneous dephasing is negligible in the doubly-excited manifold, where interaction-induced dephasing dominates, but it is included as a correction for singly-excited states. In Ref.~\cite{Covolo2025}, we have shown that this non-Markovian process can be accurately simulated by harmonically coupling a ladder of singly-excited Dicke states $\ket{\phi_n}$ with $\ket{\phi_0}=\ket{\psi_0}$, constructed through successive applications of a displacement Hamiltonian $\oH_d=\omega_0\sum_n\sqrt{n}\ket{\phi_{n-1}}\bra{\phi_n}+\text{h.c.}$. In principle, as the interaction-induced and the temperature-induced dephasing processes are uncorrelated, the asymmetric states \ket{\phi_{n>0}} should be maximally orthogonal to the states \ket{\psi_{n>0}}: $\forall (n,n')>0,\; |\langle\phi_n|\psi_{n'}\rangle|^2 \sim 1/N$. As thermal dephasing is a small correction to our dynamics, to avoid increasing the Hilbert space we form the ladder of states $\{\ket{\phi_{n>0}}\}$ via a discrete Fourier transform $\ket{\phi_n}=\sum_{n'=1}^{n_{max}}\ee^{2i\pi nn'/n_{max}}\ket{\psi_{n'}}/\sqrt{n_{max}}$, which ensures $\forall (n,n')>0,\; |\langle\phi_n|\psi_{n'}\rangle|^2= 1/n_{max}$. As shown in Ref.~\cite{Covolo2025}, the last state \ket{\phi_{n_{max}}} of the ladder decays in a singly-excited continuum \ket{c} with a rate $\gamma_T=\omega_0 (2/\pi)^{(-1)^{n_{max}}/2}n_{max}!!/(n_{max}-1)!!$.

Writing ${\bf k}=(n_c,n_d,l)$ and $n=n_c+n_d+2l$ for compactness, and defining the Lindblad superoperators $\mathcal{D}[\oA,\oB](\orho) = \oA\orho\oB^\dag-\{\oA^\dag\oB,\orho\}/2$ and $\mathcal{L}[\oA] = \mathcal{D}[\oA,\oA]$, the master equation governing the driving dynamics is then given by
\begin{align}
\notag
\dot{\orho} = \mathcal{Z}_D(\orho)
= & -i \left[ \frac{\Omega}{2}(\oS+\oSd)+\omega_0(\oa_T+\oad_T)+\hat{\Lambda}, \orho \right]\\
\notag
 &~+ 2 \gamma_r \mathcal{L} [\hat{n}_r] (\orho)
 + \sum_{{\bf k},{\bf k}'} 2 \Gamma_{{\bf k},{\bf k}'} \mathcal{D}[\ketbra{C}{\Psi_{\bf k}},\ketbra{C}{\Psi_{\bf k'}}](\orho)
 + 2\gamma_T\mathcal{L}[\ketbra{c}{\phi_{n_{max}}}](\orho),\\
\notag
\oS =&~\ketbra{G}{\psi_0}+\sum_{\bf k} S_{n;n_d,l} \ketbra{\psi_n}{\Psi_{\bf k}},\\
\notag
\oa_T =&~\sum_{n=1}^{n_{max}} \sqrt{n} \ketbra{\phi_{n-1}}{\phi_n},\\
\notag
\hat{n}_r =&~\sum_{n}\ketbra{\psi_n}{\psi_n}
 + 2\sum_{\bf k}\ketbra{\Psi_{\bf k}}{\Psi_{\bf k}}
 +\ketbra{c}{c}+2\ketbra{C}{C},\\
\label{eq:masterEqDrive}
\hat{\Lambda}-i\hat{\Gamma} =&~V_0 \left(z_0 \sum_{\bf k} \ketbra{\Psi_{\bf k}}{\Psi_{\bf k}} - \oG^{-1}\right)
\end{align}
$\oG$ being given by Eq.~\ref{eq:ResolvSpherElts}.

\subsection{R/not(R) measurement: mapping to a photon}

For the R/not(R) measurement, the master equation includes additional terms describing the mapping of the symmetric state $\ket{R}$ onto a photon and the decay of the latter out of the cavity.

The cavity mode, described by the photon annihilation operator $\oa$, is strongly coupled to symmetric excitations in the intermediate state, described by the annihilation operator $\oT=\int\dd^3{\bf r}\,\sqrt{\mu({\bf r})}\os_{ge}({\bf r})$ analogous to $\oS$, via the Hamiltonian term $g(\oad\oT+\oTd\oa)$ where $g = 2\pi \times \unit{9.70(6)}{\mega\hertz}$. Considering that the system contains two excitations at most, and writing $\ket{G}$ the collective ground state of the whole system, the conjugates of these operators generate the states $\ket{1}=\oad\ket{G}$, $\ket{2}=(\oad)^2\ket{G}/\sqrt{2}$, $\ket{e}=\oTd\ket{G}$, $\ket{E}=(\oTd)^2\ket{G}/\sqrt{2}$, $\ket{e1}=\oTd\oad\ket{G}$, $\{\ket{\psi_n e}=\oTd\ket{\psi_n}\}$, and $\{\ket{\psi_n 1}=\oad\ket{\psi_n}\}$. In this basis,
\begin{align}
\notag
\oS =&~\ketbra{G}{\psi_0}
 +\sum_{\bf k} S_{n;n_d,l} \ketbra{\psi_n}{\Psi_{\bf k}}
 + \ketbra{1}{\psi_0 1}+ \ketbra{e}{\psi_0 e},\\
\notag
\hat{n}_r =&~\sum_{n}\left(\ketbra{\psi_n}{\psi_n}+\ketbra{\psi_n e}{\psi_n e}+\ketbra{\psi_n 1}{\psi_n 1}\right) + 2\sum_{\bf k}\ketbra{\Psi_{\bf k}}{\Psi_{\bf k}} +\ketbra{c}{c}+2\ketbra{C}{C},\\
\oa =&~\ketbra{G}{1}+\ketbra{e}{e1}+\sqrt{2}\ketbra{1}{2}+\sum_{n}\ketbra{\psi_n}{\psi_n 1},\\
\oT =&~\ketbra{G}{e}+\ketbra{1}{e1}+\sqrt{2}\ketbra{e}{E}+\sum_{n}\ketbra{\psi_n}{\psi_n e}.
\end{align}
Other operators in the driving part of the master equation (Eq.~\ref{eq:masterEqDrive}) remain unchanged. In particular, we neglect the thermal dephasing of the states containing an intermediate or a photonic excitation, where the decay is dominated respectively by the atomic dipole decay rate $\gamma=2\pi\times\unit{2.87}{\mega\hertz}$ and the field decay rate $\kappa=2\pi\times\unit{2.89}{\mega\hertz}$. These decays translate into additional terms $2\gamma\mathcal{L}[\oT](\orho)+2\kappa\mathcal{L}[\oa](\orho)$ in the master equation.

The mapping laser, with a Rabi frequency $\Omega_m$ increasing from $0$ to $2 \pi \times \unit{7.4(1)}{\mega\hertz}$, converts Rydberg excitations into intermediate-state ones via the Hamiltonian term $(\Omega_m\oQ+\Omega_m^*\oQd)/2$ where $\oQ=\int\dd^3{\bf r}\os_{er}({\bf r})$ (see cyan arrows and curve in Fig.~3(c)). In principle, $\oQ$ generates the states $\oQ\ket{\psi_n}$, $\oQ\ket{\Psi_{\bf k}}/\sqrt{2}$ and 
$\oQ^2\ket{\Psi_{\bf k}}/2$. This doubles the basis size quite uselessly, these states being short-lived and weakly populated. Instead, we keep only the states defined above, where the intermediate excitation is symmetric, to describe the superatom-to-photon mapping without assuming its adiabaticity, and treat the other states generated by $\oQ$ as decay channels. For this, we rewrite $\oQ=\oTd\oS+\oQ_\perp$ where $\oTd\oS$ converts a symmetric Rydberg excitation into its intermediate counterpart, while $\oQ_\perp$ does the same for asymmetric excitations. We consider that the latter lie outside of our Hilbert space, described by a projector $\oP$, and that in absence of laser coupling their evolution is described by a non-Hermitian decay Hamiltonian $-i\gamma$. The readout term $\Omega_m\oQ/2+\text{h.c.}$ is now separated in two parts: $\Omega_m\oTd\oS/2 +\text{h.c.}$  is kept in the Hamiltonian, while $\Omega_m\oQ_\perp/2+\text{h.c.}$ is converted, using the resolvent formalism \cite{Cohen1998.ch3}, into a non-Hermitian Hamiltonian
\begin{equation}
\oP\frac{\Omega_m\oQ_\perp+\Omega_m^*\oQd_\perp}{2}\frac{1}{z+i\gamma}\frac{\Omega_m\oQ_\perp+\Omega_m^*\oQd_\perp}{2}\oP
 \approx -i\frac{|\Omega_m|^2}{4\gamma}\oP\oQd_\perp\oQ_\perp\oP
 = -i\frac{|\Omega_m|^2}{4\gamma}\left(\hat{n}_r-\oSd\oS\right)
\end{equation}
where we approximated the effective energy $z$ by $0$ for a resonant readout, and used $\oP\oQd\oQ\oP=\hat{n}_r$ and $\oP\oQd\oTd\oS\oP=\oSd\oS$. As it is purely anti-Hermitian, it translates into decay terms in the master equation, with a rate $|\Omega_m|^2/(4\gamma)$. The different decay channels should in principle correspond to spatially localized states but, for simplicity, we use our basis states and consider that the states containing two Rydberg excitations decay in \ket{c}.

The master equation governing the whole process, describing both the driving and the superatom-to-photon mapping, becomes $\dot{\orho} = \mathcal{Z}_D(\orho) + \mathcal{Z}_M(\orho)$ where
\begin{align}
\nonumber
\mathcal{Z}_M(\orho)
=&~
-i \left[ \frac{\Omega_m}{2}\oTd\oS+g\oad\oT+\text{h.c.}, \orho \right]
 + 2\gamma\mathcal{L}[\oT](\orho)
 + 2\kappa\mathcal{L}[\oa](\orho)\\
\nonumber 
&~+ 2\frac{|\Omega_m|^2}{4\gamma}\left(\sum_{n>0}
 \mathcal{L}[\ketbra{G}{\psi_n}+\ketbra{e}{\psi_n e}+\ketbra{1}{\psi_n 1}](\orho)
 + \mathcal{L}[\ketbra{G}{c}](\orho)
 + 2\mathcal{L}[\ketbra{c}{C}](\orho) \right) \\
 &~ + 2\frac{|\Omega_m|^2}{4\gamma} \sum_{{\bf k},{\bf k}'} (2\delta_{{\bf k},{\bf k}'} - \delta_{n,n'}S_{n;n_d,l}S_{n;n_d',l'}) \mathcal{D}[\ketbra{c}{\Psi_{\bf k}},\ketbra{c}{\Psi_{\bf k'}}](\orho).
\end{align}
When the driving and the mapping laser pulses are well separated in time, which is the case in our experiment (see Fig.3(c)), the asymmetric states play no role in the mapping process, and the decay terms on the second and third lines of the above equation can be neglected to speed up numerics.

The number of emitted photons is given by $n_{phot}=2\kappa_0 \int \Tr(\oad\oa\orho(t))\dd t$ where $\kappa_0=2\pi\times\unit{2.58}{\mega\hertz}$ is the measured leakage rate of the cavity field through the input/output coupler. The simulation thus accounts for losses arising from the finite cooperativity, the imperfectly adiabatic mapping, and the cavity extraction efficiency. An additional efficiency factor $\eta_{mm}=0.82$, included in the right plot in Fig.3(c), can be attributed to the imperfect mode-matching of the off-resonant drive beam $1$ and to wavefront inhomogeneities of the drive beam $2$.

\end{document}